\documentclass[12pt]{article}
\usepackage[dvipdfmx]{graphicx}
\usepackage{bmpsize}
\usepackage{bbding}

\usepackage[utf8]{inputenc}

\usepackage[square,comma,numbers,sort&compress]{natbib}
\usepackage[left=20mm,right=20mm,top=20mm,bottom=20mm]{geometry}
\usepackage{bm}

\usepackage{epsf}
\usepackage{color}
\usepackage{amsmath}
\usepackage{amssymb}
\usepackage{latexsym}
\usepackage{slashed}
\usepackage{float}
\usepackage{cases}
\usepackage{bm}
\usepackage{arydshln}
\usepackage{hyperref}
\hypersetup{ 
 setpagesize=false,
 bookmarksnumbered=true,%
 bookmarksopen=true,%
 colorlinks=true,%
 linkcolor=blue,
 citecolor=red,
}
\usepackage{comment}

\usepackage{listings}
\usepackage{xcolor}

\definecolor{codegreen}{rgb}{0,0.6,0}
\definecolor{codegray}{rgb}{0.5,0.5,0.5}
\definecolor{codepurple}{rgb}{0.58,0,0.82}
\definecolor{backcolour}{rgb}{0.95,0.95,0.92}

\lstdefinestyle{mystyle}{
    backgroundcolor=\color{backcolour},   
    commentstyle=\color{codegreen},
    keywordstyle=\color{magenta},
    numberstyle=\tiny\color{codegray},
    stringstyle=\color{codepurple},
    basicstyle=\ttfamily\footnotesize,
    breakatwhitespace=false,         
    breaklines=true,                 
    captionpos=b,                    
    keepspaces=true,                 
    numbers=left,                    
    numbersep=5pt,                  
    showspaces=false,                
    showstringspaces=false,
    showtabs=false,                  
    tabsize=2
}

\usepackage{mathtools}

\DeclarePairedDelimiterX\braket[2]{\langle}{\rangle}{#1 \delimsize\vert #2}

\usepackage{multicol}
\usepackage{lipsum}

\usepackage{braket}

\newcommand{\wt}{\widetilde}

\newcommand{\al}[1]{\begin{align}#1\end{align}}
\newcommand{\als}[1]{\begin{align*}#1\end{align*}}
\newcommand{\bp}{\begin{pmatrix}}
\newcommand{\ep}{\end{pmatrix}}
\newcommand{\bb}{\begin{bmatrix}}
\newcommand{\eb}{\end{bmatrix}}
\newcommand{\nn}{\nonumber\\}

\newcommand{\ol}{\overline}

\newcommand{\paren}[1]{\left(#1\right)}
\newcommand{\sqbr}[1]{\left[#1\right]}
\newcommand{\ab}[1]{\left|#1\right|}
\newcommand{\fn}[1]{\!\left(#1\right)}
\newcommand{\br}[1]{\left\{#1\right\}}

\newcommand{\tx}[1]{\text{#1}}
\newcommand{\wh}[1]{\widehat{#1}}
\newcommand{\ov}{\over}

\newcommand{\beq}{\begin{equation}}
\newcommand{\eeq}{\end{equation}}
\newcommand{\bea}{\begin{eqnarray}}
\newcommand{\eea}{\end{eqnarray}}

\newcommand{\pal}{\partial}

\newcommand{\fulltoday}{\number\day\space \ifcase\month\or
    January\or February\or March\or April\or May\or June\or
    July\or August\or September\or October\or November\or December\fi
    \space\number\year}

 \makeatletter
    
    \@addtoreset{equation}{section}
  \makeatother

\usepackage{calc}
\newcounter{hours}\newcounter{minutes}
\makeatletter                   
\renewcommand*{\thehours}{\two@digits\c@hours}
\renewcommand*{\theminutes}{\two@digits\c@minutes}
\makeatother

\begin{document}

\allowdisplaybreaks[2]
\renewcommand{\thefootnote}{*}

\newlength{\mylength}

\title{
%
Prospects of five-dimensional $L_{\mu} - L_{\tau}$ gauge interactions \\
in the light of elastic neutrino-electron scatterings: \\ The scope of the DUNE near detector
}

\author{
Dibyendu Chakraborty\thanks{E-mail: \tt dc282@snu.edu.in}, \mbox{}
Arindam Chatterjee\thanks{E-mail: \tt arindam.chatterjee@snu.edu.in}, \\[5pt]
Ayushi Kaushik\thanks{E-mail: \tt ak356@snu.edu.in}, \mbox{} and
Kenji Nishiwaki\thanks{E-mail: \tt kenji.nishiwaki@snu.edu.in}
\bigskip\\
\it\normalsize
Shiv Nadar Institution of Eminence, \\
\it\normalsize
Tehsil Dadri, Gautam Buddha Nagar, Uttar Pradesh,
201314, India
}
\maketitle
\begin{abstract}
\noindent
We discuss the future prospects of a minimally five-dimensional version of the well-motivated
scenario for addressing the discrepancy in the muon anomalous magnetic moment, the $U\fn{1}_{L_{\mu} - L_{\tau}}$ extension of the standard model (SM) gauge symmetry.
Here, multiple associated massive gauge bosons appear thanks to the five-dimensional $U\fn{1}_{L_{\mu} - L_{\tau}}$ gauge symmetry, and they contribute to the 
muon $(g-2)$ and also other processes. We focus on the powerful probe of elastic neutrino-electron scatterings since the upcoming DUNE experiment will explore MeV-scale uncharted regions by previous experiments (e.g., CHARM-II and Borexino) in the 
near future. We found that even with small kinetic mixing parameters, much 
of the parameter space, including those satisfying muon $(g-2)$, can be probed using 
several years of data from the DUNE experiment, focusing on the near detector. In our 
scenario, interference effects between intermediate-state gauge bosons play an important 
role. Our results include comparisons between flat and warped extra dimensions.
\end{abstract}

\newpage

\renewcommand\thefootnote{\arabic{footnote}}
\setcounter{footnote}{0}


\section{Introduction}

Over the past few decades, there has been substantial progress in 
our understanding of the lepton sector in the Standard Model~(SM). In particular, 
several important parameters in the active neutrino sector of the SM have been 
determined or constrained, which includes not only the differences between 
their masses squared and the mixing angles but also the CP-violating phase.
Also, there are continuing developments in the measurements of muon and 
electron dipole moments. Furthermore, the lepton sector has received a particularly 
large amount of attention in recent years as a portal for physics beyond the SM.

A well-motivated gauge extension of the SM is the $U\fn{1}_{L_\mu - L_\tau}$ 
symmetry, where $+1$ and $-1$ charges are assigned for the second-generation 
and third-generation leptons of the SM respectively~\cite{Foot:1990mn,He:1990pn,He:1991qd,Foot:1994vd}. Interestingly, this gauge extension is anomaly-free without the necessity to introduce right-handed neutrinos. After the 
spontaneous breakdown of the gauge symmetry, a new neutral massive gauge 
boson emerges, which can contribute to various phenomena. This kind of simple 
gauge extension (which possibly includes new particles other than 
a scalar boson that makes the $U\fn{1}_{L_\mu - L_\tau}$ boson massive through 
the Higgs mechanism) can address active-neutrino textures~\cite{Branco:1988ex,Heeck:2011wj,Asai:2017ryy,Asai:2018ocx,Asai:2019ciz,Joshipura:2019qxz,Araki:2019rmw,Fukuyama:2020swd,Bauer:2020itv,Majumdar:2020xws,Amaral:2021rzw},
dark matter candidates~\cite{Baek:2008nz,Baek:2015fea,Patra:2016shz,Biswas:2016yan,Biswas:2016yjr,Asai:2017ryy,Arcadi:2018tly,Kamada:2018zxi,Foldenauer:2018zrz,Asai:2019ciz,Okada:2019sbb,Asai:2020qlp,Holst:2021lzm,Tapadar:2021kgw,Heeck:2022znj,Nagao:2022osm,KA:2023dyz,Figueroa:2024tmn}, the muon $\paren{g-2}$~\cite{Baek:2001kca,
Ma:2001md,Harigaya:2013twa,Altmannshofer:2016brv,Hapitas:2021ilr}, the Hubble 
Tension~\cite{Escudero:2019gzq,Araki:2021xdk,Carpio:2021jhu,Asai:2023ajh}, and 
leptogenesis scenarios~\cite{Asai:2020qax,Borah:2021mri,Eijima:2023yiw,Granelli:2023egb,Wada:2024cbe}.

In general, apart from the strength of the respective gauge coupling, 
the detectability of such scenarios highly depends on the  mass of the new gauge 
boson and the strength of the kinetic mixing term between $U\fn{1}_{L_\mu - L_\tau}$ 
and $U\fn{1}_Y$, which corresponds to the hypercharge. Here, we focus our attention 
on the situation where the mass scale is set in the ballpark of MeV. 
This choice is motivated as we can accommodate the discrepancy 
in the muon $\fn{g-2}$ between the experimental results and the theoretical prediction 
in the SM, as we will mention in detail later.
Such scenarios have been probed through various physical processes, 
e.g., Neutrino Trident Production~\cite{Altmannshofer:2014pba,Altmannshofer:2019zhy,
Ballett:2019xoj,Shimomura:2020tmg}, Supernova limits~\cite{Croon:2020lrf,Cerdeno:2023kqo,Manzari:2023gkt,Akita:2023iwq,Lai:2024mse}, White Dwarf Cooling~\cite{Foldenauer:2024cdp},
Coherent Elastic Neutrino Nucleus Scattering~\cite{Abdullah:2018ykz}, and rare 
Kaon decay~\cite{Ibe:2016dir, Asai:2024pzx}. Also, { such models have been surveyed} 
in the light of various experiments, e.g., in
IceCube~\cite{Kamada:2015era,Araki:2015mya},
Belle-II~\cite{Kaneta:2016uyt,Araki:2017wyg,Jho:2019cxq,Asai:2021wzx,Bandyopadhyay:2022klg},
LHC~\cite{Nomura:2020vnk,Galon:2019owl},
$e^-e^+$ colliders~\cite{Nomura:2018yej,Zhang:2020fiu},
$e^-p$ colliders~\cite{Nomura:2018yej,Hou:2019omx},
MUonE~\cite{Asai:2021wzx},
NA64-$\paren{e,\mu}$~\cite{Gninenko:2018tlp,Krnjaic:2019rsv,Sieber:2021fue,NA64:2022rme,NA64:2024klw,Andreev:2024lps},
$\tx{M}^3$~\cite{Kahn:2018cqs},
and (other) Beam Dump Experiments~\cite{Cesarotti:2022ttv,Rella:2022len,Moroi:2022qwz,Ariga:2023fjg} (see also~\cite{Asai:2023dzs}).\footnote{
A thorough analysis based on experimental data up to $2018$ is available in~\cite{Bauer:2018onh} (see also~\cite{Patrick_Foldenauer-Dissertation}).
}

In this paper, we will focus on an extension of the vanilla $U\fn{1}_{L_\mu - L_\tau}$ 
scenario in four dimensions into five dimensions~(5D), where
one minuscule compact spatial direction is introduced.
If the $U\fn{1}_{L_\mu - L_\tau}$ gauge symmetry lives in the 5D bulk space, it is reduced to the emergence of multiple numbers of massive gauge bosons in the four-dimensional point of view, as a Kaluza-Klein~(KK) tower~\cite{Ponton:2012bi}.
Based on this idea, the origin of the MeV mass scale can be reduced to the inverse size of the compact extra dimension.
Also, the existence of the bulk extra-dimensional space suggests good connections between the SM sector and the other sectors involving particles, which contributes to physics beyond the SM.
An interesting aspect of such a scenario is multiple neutral gauge bosons interacting with the muon, which contribute to the $\paren{g-2}_\mu$.
Elastic Neutrino-Electron Scattering, represented by E$\nu$ES,
$e^- \nu_X \to e^- \nu_X$ ($X$: $e,\ol{e},\mu,\ol{\mu}$) {offers significant insight into the properties and interactions of these MeV-scale neutral gauge bosons}.\footnote{
Throughout this paper, we use the following convention for the neutrino flavours, including whether it is a particle or an anti-particle:
$\nu_{\over{X}} = \overline{\nu_X}$.
}
This channel was surveyed by the experiments CHARM-II~\cite{CHARM-II:1993phx,CHARM-II:1994dzw}, TEXONO~\cite{TEXONO:2009knm,Wong:2015kgl}, and GEMMA~\cite{Beda:2009kx,Beda:2010hk} and their constraints were imposed.
Also, the Borexino experiment has investigated this channel by use of solar neutrinos~\cite{Bellini:2011rx,Kumaran:2021lvv}.
Recently, it has been recognised that the DUNE experiment~\cite{DUNE:2016hlj,DUNE:2020ypp}, which is a next-generation high-precision neutrino oscillation experiment,
also holds the powerful potential to probe new physics through E$\nu$ES by focusing on the near detector~\cite{Bakhti:2018avv,DeRomeri:2019kic,Ballett:2019bgd,Coloma:2020lgy,Kelly:2020dda,Breitbach:2021gvv,Dev:2021qjj,Chakraborty:2021apc,Chauhan:2022iuh,Asai:2022zxw,Melas:2023olz,Singh:2023nek,Felkl:2023nan,Candela:2024ljb}; see also~\cite{Bertuzzo:2021opb}.
In this paper, we will provide a dedicated analysis of current constraints and future 
prospects of the 5D vanilla $U\fn{1}_{L_\mu - L_\tau}$ scenarios through the E$\nu$ES 
processes, where both the flat and warped 5D geometries are considered.

The paper is organised as follows.
In Section~\ref{sec:Setup}, we { describe the model under consideration
in five dimensions and provide the basic formulas for analysis.}
In Section~\ref{sec:Experiments}, some details of experiments measuring E$\nu$ES 
processes are given.
In Section~\ref{sec:Results}, we discuss current constraints on our scnario and 
its future discovery potential in the DUNE experiment.
Finally, in Section~\ref{sec:Conclusions}, we discuss the future prospects and conclude.
In the appendices~\ref{sec:KK-decomposition}, \ref{sec:kinematics}, and \ref{sec:supplemental-plots}, the details of the KK decompositions, the kinematics of E$\nu$ES (in the laboratory frame), and useful supplemental plots have been provided, respectively.

\section{Model Setups and Basic Formulas
\label{sec:Setup}}

\begin{figure}[t]
\centering
\includegraphics[width=0.35\textwidth]{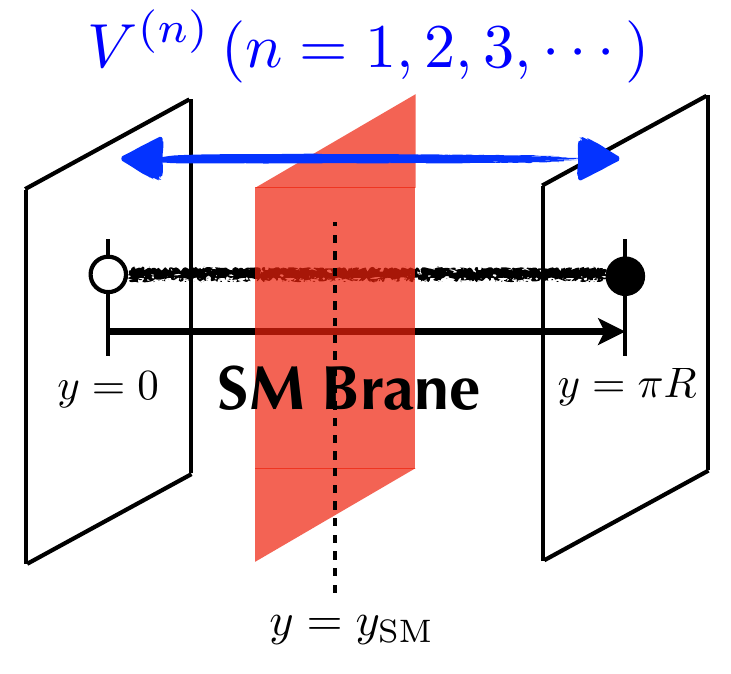}
\caption{
A schematic picture of our 5D setup.
All of the SM particles are confined within the zero-thickness brane located at $y=y_\tx{SM}$.
Only the KK $U\fn{1}_{L_\mu - L_\tau}$ gauge bosons propagate in the bulk space.
The white and black dots indicate that the Neumann and Dirichlet boundary conditions are taken for the four-dimensional vector part of the 5D gauge boson, respectively.
}
\label{fig:5D-setup}
\end{figure}

In this part, we will introduce a brief description of the five-dimensional $U\fn{1}_{L_\mu - L_\tau}$ scenario
on a compact extra spatial dimension, where all of the SM particles are confined on a {three spatial dimensional} brane located at $y = y_\tx{SM}$, while the
$U\fn{1}_{L_\mu - L_\tau}$ gauge boson and the respective KK particles live in the five-dimensional bulk.
We adopt the coordinate for the extra dimension, $y\!\!:\!\sqbr{0,\pi R}$,
where $R$ is the `radius' of the compact extra dimension.
A schematic picture of our setup is shown in Fig.~\ref{fig:5D-setup}.\footnote{
A similar setup was considered in~\cite{Rizzo:2018ntg}, see also earlier works~\cite{McDonald:2010iq,McDonald:2010fe,Jaeckel:2014eba}.
Recent associated papers are found~\cite{Rizzo:2018joy,Landim:2019epv,Brax:2019koq,Landim:2019ufg,Landim:2019gdj,Bernal:2020fvw,Garbrecht:2020tng,Rizzo:2020ybl};
those in stringy {contexts}~\cite{Anchordoqui:2020tlp,Anchordoqui:2021lmm,Antoniadis:2021mqz,Cheung:2022kjd,Anastasopoulos:2022wob}, which include effective operator analysis of such multiple gauge bosons of constraints from LEP, LHC and forward physics factory~\cite{Anchordoqui:2021lmm,Cheung:2022kjd}.
See also discussions on the contribution to $\paren{g-2}_\mu$ of anomalous $U\paren{1}$ gauge bosons~\cite{Anastasopoulos:2022ywj,Anastasopoulos:2024bxx}.
}
We note that the generalisation of the position of the SM brane (which can be located anywhere in the extra dimension, including the endpoints) has just been introduced for our phenomenological purpose.

{Note that the purpose of this paper is a phenomenological investigation of current constraints and future prospects for the 5D scenario through E$\nu$ES processes, as we will see later manifestly, where the profile of the bulk wavefunction of the KK gauge bosons is involved in determining the magnitude of such processes.
Both the flat and the warped cases (without addressing the gauge hierarchy problem) are of phenomenological interest; we have studied both of these scenarios using suitable phenomenological parametrisation.
Here, the profiles of the wavefunctions are different in the two cases, so as we will see later, even if we take the same common parameters, the final results will be different.
}
We do not focus on potential problems in the stability of such a system in this paper, particularly for the warped background~\cite{Goldberger:1999uk,Goldberger:1999un}.\footnote{
{Here, we comment on the coordination such that matter is localised outside the endpoints of the entire extra dimension, as shown in Fig.~\ref{fig:5D-setup}. In the flat background, since the solution of the Einstein equations in the extra-dimensional direction is trivial, terms localised at a point in the extra dimension can be introduced by hand without contradiction.
On the other hand, in the warped background, since the solution of the Einstein equations in the extra-dimensional direction is non-trivial, it is debatable whether the system is stable under the introduction of such a localisation point.
One concrete way of realising this is, for example,~\cite{Lee:2021wau}.}
}

Here, we assign the twisted boundary conditions for the 5D gauge boson, where the Neumann and Dirichlet boundary conditions are adopted
at $y=0$ and $y=\pi R$, respectively,
and no massless zero mode emerges.
In other words, we remove the massless mode without relying on the Higgs mechanism by an extra singlet scalar.

\subsection{Effective four-dimensional Lagrangian via KK decomposition
\label{sec:effective-Lagrangian}}

Here, the necessary setup for phenomenological calculations is introduced in terms of effective field theory in order to provide a common tool
treating the flat and warped backgrounds together.
Details of the KK decompositions are available in Appendix~\ref{sec:KK-decomposition}.

The effective four-dimensional (4D) form of the 4D vectors' free part reads
\al{
{\cal L}_\tx{eff}^{\paren{V\tx{-free}}}
	&=
		\sum_n
		\br{
			-{1\ov 4} \wh{V}_{\mu\nu}^{\paren{n}} \wh{V}^{\paren{n}\mu\nu}
			+{\epsilon_n\ov 2c_W} \wh{V}_{\mu\nu}^{\paren{n}} \wh{B}^{\mu\nu}
			+{1\ov 2} M_n^2 \wh{V}_\mu^{\paren{n}} \wh{V}^{\paren{n}\mu}
		} \nn
	&\quad
		-{1\ov 4} \wh{B}_{\mu\nu} \wh{B}^{\mu\nu} -{1\ov 4} {W}^3_{\mu\nu} {W}^{3\mu\nu}
		+{1\ov 2} \paren{ v g_2 W^3_\mu + v g_1 \wh{B}_\mu \ov 2}^2,
	\label{eq:effectiveLagrangian-Vfree}
}
where $n = 1,2,3,\cdots$ discriminate massive KK states;
{ $\wh{V}^{\paren{n}}_\mu$ denotes the 4D gauge eigenstate field
of the $n$-th KK excitation of the $U\fn{1}_{L_\mu - L_\tau}$ gauge boson, $\wh{B}_\mu$ 
and $W^3_\mu$ are gauge-eigenstate 4D gauge fields of the 
$U(1)_Y$}, and the 3rd component of $SU(2)_W$, respectively,
where each field strength is defined as usual, $V_{\mu\nu} \leftrightarrow \pal_\mu V_\nu - \pal_\nu V_\mu$, namely for
$\wh{V}^{\paren{n}}_\mu$, $\wh{B}_\mu$ and $W^3_\mu$ as follows,
\al{
\wh{V}_{\mu\nu}^{\paren{n}} &:= \pal_\mu \wh{V}^{\paren{n}}_\nu - \pal_\nu \wh{V}^{\paren{n}}_\mu,&
\wh{B}_{\mu\nu} &:= \pal_\mu \wh{B}_\nu - \pal_\nu \wh{B}_\mu,&
{W}^3_{\mu\nu} &:= \pal_\mu {W}^3_\nu - \pal_\nu {W}^3_\mu.&
}
$\epsilon_n$ is the effective kinetic mixing factor between $\wh{V}^{\paren{n}}_\mu$ and $\wh{B}_\mu$.
$M_n$ is the $n$-th KK mass.
Note that $\epsilon_n$ and $M_n$ take different forms in the cases of the flat and warped backgrounds.
$v \simeq 246\,\tx{GeV}$ is the Higgs vacuum expectation value~(VEV).
$c_W$, $s_W$, $g_2$ and $g_1$ are the cosine and the sine of the Weinberg angle, the $SU(2)_W$ gauge coupling, and the $U(1)_Y$ gauge coupling.\footnote{
We have adopted the SM convention on the review~\cite{Denner:1991kt}.
Note that the scheme of connection between two different $U(1)$ gauge bosons through a kinetic mixing term was considered in~\cite{Holdom:1985ag},
see also some earlier associated works~\cite{Holdom:1986eq,DelAguila:1993px,Dienes:1996zr,Babu:1996vt,Rizzo:1998ut}.
We have referred to the discussions in~\cite{Rizzo:2018ntg} and also in~\cite{Curtin:2014cca}.
}

The introduction { of} the secondary basis,
\al{
\wh{B}_\mu 
	&=
		B_\mu + \sum_n {\epsilon_n \ov c_W} \wt{V}^{\paren{n}}_\mu, \\
\wh{V}^{\paren{n}}_\mu
	&=
		\wt{V}^{\paren{n}}_\mu, \\
\bp \wt{Z}_\mu \\ A_\mu \ep
	&=
		\bp c_W & s_W \\ -s_W & c_W \ep
		\bp W^3_\mu \\ B_\mu \ep,
}
brings us into
\al{
{ {\cal L}}_\tx{eff}^{\paren{V\tx{-free}}}
	&=
		-{1\ov 4} \sum_n \wt{V}_{\mu\nu}^{\paren{n}} \wt{V}^{\paren{n}\mu\nu}
		-{1\ov 4} \wt{Z}_{\mu\nu} \wt{Z}^{\mu\nu}
		-{1\ov 4} A_{\mu\nu} A^{\mu\nu} \nn
	&\quad
		+{1\ov 2} \sum_n M_n^2 \wt{V}_\mu^{\paren{n}} \wt{V}^{\paren{n}\mu}
		+{1\ov 2} m_{Z,0}^2 \wt{Z}_\mu \wt{Z}^\mu
		+ m_{Z,0}^2  t_W \sum_n\epsilon_n \wt{Z}_\mu \wt{V}^{\paren{n}\mu}
		+{1\ov 2} m_{Z,0}^2 t_W^2 \sum_{n,n'} \epsilon_n \epsilon_{n'} \wt{V}^{\paren{n}}_\mu \wt{V}^{\paren{n'}\mu},
	\label{eq:S_eff-Vfree-1}
}
where $A_\mu$ is the photon field, $t_W = s_W/c_W$, and $m_{Z,0}^2 := \paren{g_1^2+g_2^2}v^2/2$.
Assuming that all of the dimensionless parameters $\br{ \epsilon_n }_{n=1,2,3,\cdots}$ are sufficiently small,
the unitary transformation diagonalises the mass terms of Eq.~\eqref{eq:S_eff-Vfree-1} are approximately given by (up to ${\cal O}\fn{\epsilon_n^2}$),
\al{
\wt{V}^{\paren{n}}_\mu
	&=
		V^{\paren{n}}_\mu - t_W {\epsilon_n m_{Z,0}^2 \ov M_n^2 - m_{Z,0}^2} Z_\mu + {\cal O}\fn{\epsilon_n^2}, \\
\wt{Z}_\mu
	&=
		Z_\mu + t_W \sum_n {\epsilon_n m_{Z,0}^2 \ov M_n^2 - m_{Z,0}^2} V^{\paren{n}}_\mu + {\cal O}\fn{\epsilon_n^2},
}
where the diagonalised form is given as\footnote{
If we take into account the 2nd-order perturbation for the eigenvalues,
the 1st-order mass eigenvalues are slightly shifted as follows:
\al{
m_{Z,0}^2
	&\to
		m_{Z,0}^2 \sqbr{ 1 - t_W^2 m_{Z,0}^2 \sum_n {\epsilon_n^2 \ov M_n^2 - m_{Z,0}^2} },&
M_n^2
	&\to
		M_n^2 \sqbr{ 1 + t_W^2 {\epsilon_n^2 m_{Z,0}^2 \ov M_n^2 - m_{Z,0}^2}  }.
	\label{eq:masssq-perturbed}
}
}
\al{
{\cal L}_\tx{eff}^{\paren{V\tx{-free}}}
	&=
		-{1\ov 4} \sum_n {V}_{\mu\nu}^{\paren{n}} {V}^{\paren{n}\mu\nu}
		-{1\ov 4} {Z}_{\mu\nu} {Z}^{\mu\nu}
		-{1\ov 4} A_{\mu\nu} A^{\mu\nu}
		+{1\ov 2} m_{Z,0}^2 Z_\mu Z^\mu\
		+{1\ov 2} \sum_n M_n^2 {V}_\mu^{\paren{n}} {V}^{\paren{n}\mu}
		+{\cal O}\fn{\epsilon_n^2},
	\label{eq:L_eff_diagonalised}
}
where $Z_\mu$ and $V_\mu^{\paren{n}}$ are the $Z$ boson and the $n$-th $U(1)_{L_\mu-L_\tau}$ gauge boson in the mass eigenbasis, respectively.
Here, we summarise the relations between the gauge eigenstates and the mass eigenstates below,
\al{
\wh{B}_\mu
	&=
		s_W Z_\mu + c_W A_\mu + \sum_n \sqbr{ {\epsilon_n\ov c_W} + s_W t_W {\epsilon_n m_{Z,0}^2\ov M_n^2-m_{Z,0}^2} }
			V^{\paren{n}}_\mu + {\cal O}\fn{\epsilon_n^2}, \\
W^3_\mu
	&=
		c_W Z_\mu - s_W A_{\mu} + c_W t_W \sum_n {\epsilon_n m_{Z,0}^2 \ov M_n^2 - m_{Z,0}^2} V^{\paren{n}}_\mu
		+ {\cal O}\fn{\epsilon_n^2}, \\
\wh{V}^{\paren{n}}_\mu
	&=
		\wt{V}^{\paren{n}}_\mu
		=
		V^{\paren{n}}_\mu - t_W {\epsilon_n m_{Z,0}^2 \ov M_n^2 - m_{Z,0}^2} Z_\mu + {\cal O}\fn{\epsilon_n^2}.
}

The effective interactions of the charged leptons and left-handed neutrinos, where we treat the neutrinos as massless
and no right-handed neutrinos are introduced, are given as follows:
\al{
{{\cal L}_\tx{eff,int}}
	&=
		\sum_{x=e,\mu,\tau}
		\sqbr{
		\ol{L'}^{x}_{L} i \gamma^\mu D_\mu {L'}^{x}_L + \ol{l'}^{x}_{R} i \gamma^\mu D_\mu {l'}^{x}_R
		},
		\label{eq:effectiveLagrangian-fermion}
}
where $L^{'x}_L = \paren{\nu^{'x}_L,\,  l^{'x}_L}^\tx{T}$ is the $x$-th $SU(2)_W$ lepton doublet in the gauge eigenstate
and $l^{'x}_R$ is the $x$-th $SU(2)_W$ charged lepton singlets. The prime symbol represents the gauge eigenstate for fermions.
The effective forms of covariant derivatives yield
\al{
D_\mu \Big|_{L^{'x}_L}
	&=
		\pal_\mu I_2 - ig_2 {\sigma^i \ov 2} W^i_\mu + ig_1 {Y_{L_L} \ov 2} I_2 \wh{B}_\mu 
		- i g' Q^x_{L_\mu-L_\tau} I_2 \sum_n \wh{V}^{\paren{n}}_\mu f_n, \\
D_\mu \Big|_{l^{'x}_R}
	&=
		\pal_\mu + ig_1 {Y_{l_R} \ov 2} \wh{B}_\mu 
		- i g' Q^x_{L_\mu-L_\tau} \sum_n \wh{V}^{\paren{n}}_\mu f_n,
}
where $\sigma^i$ is the $i$-th Pauli matrix, $ I_2$ represents the two-dimensional identity matrix,
the hypercharges $Y_{L_L}$ and $Y_{l_R}$ take $-1$ and $-2$, respectively,
$g'$ is the effective 4D gauge coupling for $U(1)_{L_\mu-L_\tau}$.
Note that $f_n$ represents the effective vertex factor for $\wh{V}^{\paren{n}}_\mu$, where it also takes different forms in flat and warped backgrounds.
The $U(1)_{L_\mu-L_\tau}$ charge for each field is taken as
\al{
Q_{L_\mu-L_\tau}\fn{L^{'e}_L} &= 0,&
Q_{L_\mu-L_\tau}\fn{L^{'\mu}_L} &= +1,&
Q_{L_\mu-L_\tau}\fn{L^{'\tau}_L} &= -1,& \nn
Q_{L_\mu-L_\tau}\fn{l^{'e}_R} &= 0,&
Q_{L_\mu-L_\tau}\fn{l^{'\mu}_R} &= +1,&
Q_{L_\mu-L_\tau}\fn{l^{'\tau}_R} &= -1.& 
}
In the mass eigenstates (without the prime symbol for leptons), the effective Lagrangian reads
\al{
&{{\cal L}_\tx{eff,int}} \nn
	&=
		\sum_{a=e,\mu,\tau} \Bigg\{
		\ol{l}^a_R i \gamma^\mu \pal_\mu l^a_R
		+ g_1 s_W \ol{l}^a_R \gamma^\mu l^a_R { Z_{\mu}}
		+ e \, \ol{l}^a_R \gamma^\mu l^a_R  { A_{\mu}}\nn
	&\quad
		+ \ol{l}^a_R \gamma^\mu g_1 \sum_n
			\sqbr{ {\epsilon_n \ov c_W} + s_W t_W { \epsilon_n m_{Z,0}^2 \ov M_n^2 - m_{Z,0}^2 } } V^{\paren{n}}_\mu l^a_R
		+ \ol{l}^a_R \gamma^\mu g' Q_{L_\mu-L_\tau}^{a} \sum_n {f_n}
			\sqbr{ V^{\paren{n}}_\mu - t_W { \epsilon_n m_{Z,0}^2 \ov M_n^2 - m_{Z,0}^2 } Z_\mu }  l^a_R \nn
	&\quad
		+ \ol{\nu}^a_L i\gamma^\mu \pal_\mu \nu^a_L
		+ \ol{l}^a_L i\gamma^\mu \pal_\mu l^a_L
		+ {g_2\ov \sqrt{2}} \paren{ \ol{\nu}^a_L \gamma^\mu l^a_L W^+_\mu + \ol{l}^a_L \gamma^\mu \nu^a_L W^-_\mu } \nn
	&\quad
		+ \ol{\nu}^a_L
		\bigg\{
			{1\ov 2} \sqbr{ \paren{g_2c_W + g_1s_W} Z_\mu 
			+ \sum_n \paren{ \paren{ g_2c_W + g_1s_W } t_W {\epsilon_n m_{Z,0}^2 \ov M_n^2 - m_{Z,0}^2} 
			+ g_1 {\epsilon_n \ov c_W} } V^{\paren{n}}_\mu } \nn
	&\phantom{\quad + \ol{\nu}^a_L \bigg\{}
		+ g' Q_{L_\mu-L_\tau}^{a} \sum_n f_n \paren{ V^{\paren{n}}_\mu - t_W { \epsilon_n m_{Z,0}^2 \ov M_n^2 - m_{Z,0}^2 } Z_\mu }
		\bigg\} \nu^a_L \nn
	&\quad
		+ \ol{l}^a_L
		\bigg\{
			e A_\mu
			+ {1\ov 2} \sqbr{ \paren{-g_2c_W + g_1s_W} Z_\mu 
			+ \sum_n \paren{ \paren{ -g_2c_W + g_1s_W } t_W {\epsilon_n m_{Z,0}^2 \ov M_n^2 - m_{Z,0}^2} 
			+ g_1 {\epsilon_n \ov c_W} } V^{\paren{n}}_\mu } \nn
	&\phantom{\quad + \ol{\nu}^a_L \bigg\{}
		+ g' Q_{L_\mu-L_\tau}^{a} \sum_n f_n \paren{ V^{\paren{n}}_\mu - t_W { \epsilon_n m_{Z,0}^2 \ov M_n^2 - m_{Z,0}^2 } Z_\mu }
		\bigg\} l^a_L 
	\Bigg\} + {\cal O}\fn{\epsilon_n^2},
	\label{eq:S_eff-Vfree-2}
}
where we introduced the shorthand notation:
\al{
Q_{L_\mu-L_\tau}^{a}
	&=
		\begin{cases}
		0 & \tx{for} \ a = e, \\
		+1 & \tx{for} \ a = \mu, \\
		-1 & \tx{for} \ a = \tau.
		\end{cases}
}
The Feynman rules for effective vertices are constructed from Eq.~\eqref{eq:S_eff-Vfree-2} straightforwardly.

\subsection{Description of the effective parameters
\label{sec:effective-parameters}}

Here, we briefly summarise the concrete description of the effective parameters appearing in Eq.~\eqref{eq:S_eff-Vfree-2}, namely,
$g'$, $f_n$, $\epsilon_n$, and $M_n$ ($n = 1,2,3,\cdots$). See Appendix~\ref{sec:KK-decomposition} for more information.

\begin{itemize}
\item
As shown in Eqs.~\eqref{eq:5Daction-flat} and \eqref{eq:5Daction-warped},
the 5D $U(1)_{L_\mu-L_\tau}$ gauge coupling $g'_\tx{5D}$ is not dimensionless and has the mass dimension $-1/2$.
In 4D effective theories, it is always accompanied with $\sqrt{\pi R}$ as the combination
\al{
g'
	&=
		{g'_\tx{5D} \ov \sqrt{\pi R}},
}
and we can consider $g'$ as a fundamental parameter.
\item
The KK decompositions of the actions in Eqs.~\eqref{eq:5Daction-flat} and \eqref{eq:5Daction-warped}
under the parametrisation~\eqref{eq:KK-expansion-form} tells us the correspondences,
\al{
\epsilon_n 
	&=
		\epsilon_4 f_V^{\paren{n}}\fn{y_\tx{SM}},&
f_n
	&=
		f_V^{\paren{n}}\fn{y_\tx{SM}},&
}
where the effective dimensionless parameter $\epsilon_4$ is defined as $\epsilon_4 := {\epsilon_\tx{5D}/\sqrt{\pi R} }$.
$\epsilon_\tx{5D}$ is an original parameter in the actions in Eqs.~\eqref{eq:5Daction-flat} and \eqref{eq:5Daction-warped}
describing the kinetic mixing effect, which has the mass dimension $-1/2$.
$f_V^{\paren{n}}\fn{y_\tx{SM}}$ represents the value of the bulk wave function of the $n$-th KK $U(1)_{L_\mu-L_\tau}$ gauge boson
at the position $y = y_\tx{SM}$ (see Fig.~\ref{fig:5D-setup}).
\item
For the flat case, { an analytical form can be obtained} for $M_n$ ($n=1,2,3,\cdots$),
\al{
M_n 
	&=
		\paren{n \,{-}\, \frac{1}{2}}\frac{1}{R}
		=
		\paren{2n-1} m_\tx{KK},
}
where we define the KK mass as $m_\tx{KK} := M_1$ as a typical mass scale.
$m_\tx{KK}$ is a physical parameter of the effective theories.
\item
For the warped case, no analytical form for $M_n$ is available. However, the implicit form is possible,
\al{
M_n = \lambda_n k = \sqbr{\lambda_n \ov \lambda_1} m_\tx{KK},
}
where we also define the KK mass as $m_\tx{KK} := M_1$ as a typical mass scale.
$k$ is an extra parameter only for the warped case with mass dimension one, which describes the magnitude of the warping toward the $y$ direction.
The dimensionless factor $\lambda_n$ can only be calculable numerically as the $n$-th root of the equation~\eqref{eq:warped_twisted-KKmass_condition}.
\end{itemize}

\subsection{Differential cross section of $e^- \nu_X \to e^- \nu_X$
\label{sec:diff-cross-section}}

\begin{figure}[t]
\centering
\includegraphics[width=0.9\textwidth]{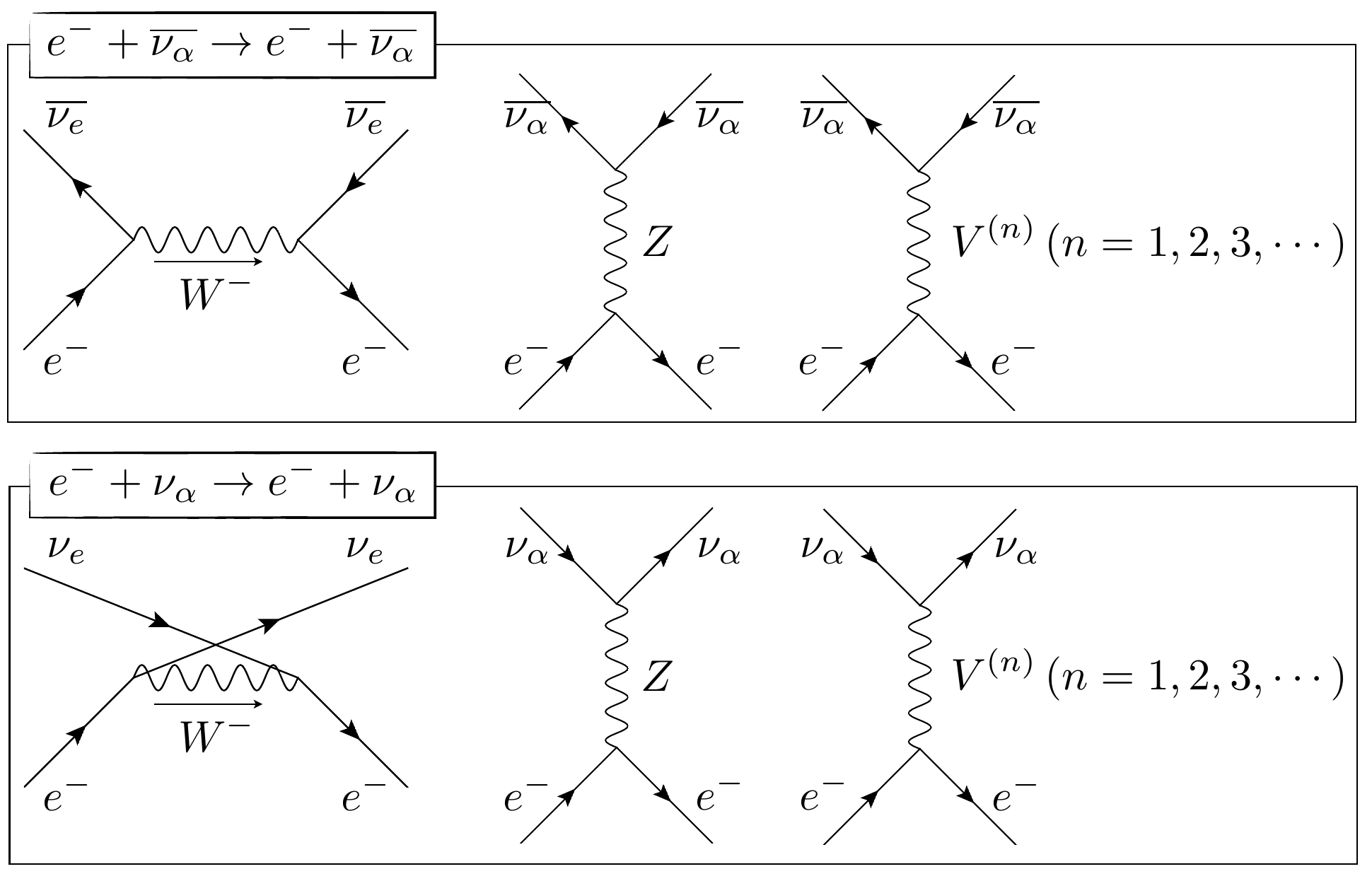}
\caption{
The Feynman diagrams contributing to the processes $e^- + \ol{\nu_\alpha} \to e^- + \ol{\nu_\alpha}$ and $e^- + {\nu_\alpha} \to e^- + {\nu_\alpha}$ ($\alpha = e,\mu,\tau$) are illustrated.
Note that the $W$-boson exchange is viable only for $\alpha = e$.
}
\label{fig:EnuES-diagrams}
\end{figure}

We summarise a general formula for the differential cross section of E$\nu$ES by following Ref.~\cite{Lindner:2018kjo},
where the unitary gauge is taken, and the contributions from the longitudinal part of the propagating gauge bosons are discarded
(see Appendix~\ref{sec:KK-decomposition} for details).
We follow the momentum convention,
\al{
\nu_\alpha\fn{p_\nu} + e^-\fn{p_e}
	&\to
		\nu_\alpha\fn{k_\nu} + e^-\fn{k_e},&
\ol{\nu_\alpha} \fn{p_\nu} + e^-\fn{p_e}
	&\to
		\ol{\nu_\alpha} \fn{k_\nu} + e^-\fn{k_e},&		
}
where $\alpha = e,\,\mu,$ or $\tau$ and the neutrino is a massless particle.
A summary of relevant Feynman diagrams is { shown} in Fig.~\ref{fig:EnuES-diagrams}.
For $e^- \ol{\nu_\alpha} \to e^- \ol{\nu_\alpha}$, for each Feynman diagram $j$, the total amplitude is denoted as\footnote{
For the electron neutrino, the $W$-boson exchange should be taken into account.
Thanks to the Fierz identity, we can deform the amplitude into the form of Eq.~\eqref{eq:amplitude_j}.
See, e.g., Eqs.~(1) and (2) of Ref.~\cite{Nishi:2004st} for details; see also Ref.~\cite{Nieves:2003in}.
}
\al{
i {\cal M}^{ss',rr'}_{e^-\ol{\nu_\alpha}}
	:=
		\sum_{j\tx{: diagrams}}
		i {\cal M}^{ss',rr'}_{e^-\ol{\nu_\alpha}, \,j},
}
with
\al{
{\cal M}^{ss',rr'}_{e^-\ol{\nu_\alpha}, \,j}
	:=
		\sqbr{ \ol{v}^s\fn{p_\nu} P_R \, \Gamma^\mu_j v^{s'}\fn{k_\nu} }
		\sqbr{ \ol{u}^{r'}\fn{k_e} \paren{\wt{\Gamma}_j}_\mu u^{r}\fn{p_e} },
	\label{eq:amplitude_j}
}
where $s$, $s'$, $r$ and $r'$ represent helicities.
Here, the effective vertex factors are introduced,
\al{
\Gamma^\mu_j
	&:=
		\gamma^\mu \paren{ c_j + d_j \gamma_5}, \nn
\wt{\Gamma}^\mu_j
	&:=
		\gamma^\mu \paren{ \wt{c}_j + \wt{d}_j \gamma_5},
	\label{eq:effective-Gamma}	
}
where $\wt{\Gamma}^\mu_j$ contains, besides the relevant vertex 
factors, the contribution from the propagator of the corresponding gauge-boson ($V_j$) exchange,
\al{
\chi_{j}^{-1}
	:=
		\frac{1}{p^2_{V_j} - m^2_{V_j} }.
}
The helicity-averaged amplitude squared is defined as
\al{
\langle \ab{{\cal M}_{e^-\ol{\nu_\alpha}}}^2 \rangle_\tx{helicity}
	:=
		\sum_{s,s'} \frac{1}{2} \sum_{r,r'} \ab{{\cal M}^{ss',rr'}_{e^-\ol{\nu_\alpha}}}^2.
}
With the help of Eq.~\eqref{eq:dsigma-ov-dTe}, the final form for $e^- \ol{\nu_\alpha} \to e^- \ol{\nu_\alpha}$
can be written down in a systematic way:
\al{
\frac{d \sigma_{e^- \ol{\nu_\alpha}}}{d T_e}
	=
		\frac{m_e}{4\pi}
		\sqbr{
			G_+^2 + G_-^2 \paren{1 - \frac{T_e}{E_\nu}}^2 - G_+ G_- \frac{m_e T_e}{E_\nu^2}
		},
	\label{eq:diff-cross-section-1}
}
where $T_e$ represents the kinetic energy of the final-state electron.
Please refer to Appendix~\ref{sec:kinematics} for details of the kinematics.
In the above equation $G_{\pm}$ describe the gauge-boson exchanges,
\al{
G_\pm
	&:=
		\sum_{j\tx{: diagrams}}
		\paren{c_j-d_j}\paren{\wt{c}_j\pm\wt{d}_j}.
	\label{eq:G-pm-factor}
}
The form for $e^- {\nu}_\alpha \to e^- {\nu}_\alpha$ can be calculated similarly:
\al{
\frac{d \sigma_{e^- {\nu}_\alpha}}{d T_e}
	=
		\frac{m_e}{4\pi}
		\sqbr{
			G_-^2 + G_+^2 \paren{1 - \frac{T_e}{E_\nu}}^2 - G_+ G_- \frac{m_e T_e}{E_\nu^2}
		},
	\label{eq:diff-cross-section-2}
}
where the two coefficients $G_+$ and $G_-$ are exchanged.

We have confirmed the coefficients for the SM in Ref.~\cite{Lindner:2018kjo}:
\al{
G_+^\tx{SM}
	&=
		-2\sqrt{2} G_F s_W^2, \\
G_-^\tx{SM}
	&=
		\begin{cases}
			\sqrt{2} G_F \paren{1-2s_W^2-2} = -2\sqrt{2} G_F \paren{s_W^2 + \frac{1}{2}}, & \tx{for } \alpha = e, \\
			\sqrt{2} G_F \paren{1-2s_W^2} = -2\sqrt{2} G_F \paren{s_W^2 - \frac{1}{2}}, & \tx{for } \alpha = \mu,\, \tau,
		\end{cases}
}
where $G_F$ represents the Fermi constant.
Note that these are consistent with Eq.~(5) and Table~II of Ref.~\cite{Chakraborty:2021apc}.

From Eq.~\eqref{eq:S_eff-Vfree-2}, it is straightforward to derive the corresponding forms of
the $c_j$, $d_j$, $\wt{c}_j$ and $\wt{d}_j$ factors defined in Eq.~\eqref{eq:effective-Gamma} for representing
the processes $e^- \ol{\nu_\alpha} \to e^- \ol{\nu}_\alpha$ and $e^- {\nu_\alpha} \to e^- {\nu}_\alpha$ ($\alpha = e,\mu,\tau$):
\begin{itemize}
\item For $W$-exchange (only for $\alpha = e$):
\al{
c^{e}_W &:= {g_2 \ov 2\sqrt{2}},&
d^{e}_W &:= -{g_2 \ov 2\sqrt{2}},&
\wt{c}^{e}_W &:= {g_2 \ov 2\sqrt{2} \chi_W},&
\wt{d}^{e}_W &:= -{g_2 \ov 2\sqrt{2} \chi_W},&
}
with
\al{
\chi_W
	:=
		\begin{cases}
		\paren{p_\nu + p_e}^2 - m_W^2 & \tx{for} \ \  e^- {\ol{\nu_e}} \to e^- {\ol{\nu_e}}, \\
		\paren{p_e - k_\nu}^2 - m_W^2 & \tx{for} \ \  e^- {\nu_e} \to e^- {\nu_e}.
		\end{cases}
}
\item For $Z$-exchange:
\al{
c_Z^{\alpha}
	&:=
		+
		{1\ov 2} \br{ {1\ov 2} \paren{g_2c_W+g_1s_W} - g'Q^\alpha_{L_\mu-L_\tau} t_W 
			\sum_n {f_n \epsilon_n m_{Z,0}^2 \ov M_n^2 - m_{Z,0}^2} }, \\
d_Z^{\alpha}
	&:=
		-
		{1\ov 2} \br{ {1\ov 2}\paren{g_2c_W+g_1s_W} - g'Q^\alpha_{L_\mu-L_\tau} t_W 
			\sum_n {f_n \epsilon_n m_{Z,0}^2 \ov M_n^2 - m_{Z,0}^2} }, \\
\wt{c}_Z^{{\alpha}}
	&:=
		{1\ov 2\chi_Z}
		\br{
			g_1s_W - 2g'Q^{{\alpha}}_{L_\mu-L_\tau}t_W \sum_n {f_n \epsilon_n m_{Z,0}^2 \ov M_n^2 - m_{Z,0}^2}
			+{1\ov 2}\paren{-g_2c_W+g_1s_W}
		}, \\
\wt{d}_Z^{{\alpha}}
	&:=
		{1\ov 2\chi_Z}
		\br{
			g_1s_W - {1\ov 2}\paren{-g_2c_W+g_1s_W}
		},
}
with
\al{
\chi_Z
	:=
		\paren{p_\nu - k_\nu}^2 - (m_Z^\tx{phys})^2,
}
where $m_Z^\tx{phys}$ represents the physical mass of the $Z$ boson.
\item For $V^{\paren{n}}$-exchange:
\al{
c_{V^{\paren{n}}}^{\alpha}
	&:=
		+
		{1\ov 2}
		\br{
			{1\ov 2} \sqbr{ \paren{g_2c_W+g_1s_W} t_W {\epsilon_n m_{Z,0}^2 \ov M_n^2 - m_{Z,0}^2} + g_1 {\epsilon_n \ov c_W} }
			+ f_n g' Q^\alpha_{L_\mu-L_\tau}
		}, \\
d_{V^{\paren{n}}}^{\alpha}
	&:=
		-
		{1\ov 2}
		\br{
			{1\ov 2} \sqbr{ \paren{g_2c_W+g_1s_W} t_W {\epsilon_n m_{Z,0}^2 \ov M_n^2 - m_{Z,0}^2} + g_1 {\epsilon_n \ov c_W} }
			+ f_n g' Q^\alpha_{L_\mu-L_\tau}
		}, \\
\wt{c}_{V^{\paren{n}}}^{{\alpha}}
	&:=
		{1\ov 2\chi_{V^{\paren{n}}}}
		\Bigg\{ {3 \ov 2}
			g_1 \paren{ {\epsilon_n\ov c_W} + s_Wt_W {\epsilon_nm_{Z,0}^2 \ov M_n^2-m_{Z,0}^2} }
			+ 2 f_n g' Q^{{\alpha}}_{L_\mu-L_\tau}
			\nn
	&\quad 
			-{1\ov 2} \paren{ g_2s_W  {\epsilon_n m_{Z,0}^2 \ov M_n^2-m_{Z,0}^2} }
		\Bigg\}, \\
\wt{d}_{V^{\paren{n}}}^{{\alpha}}
	&:=
		{1\ov 2\chi_{V^{\paren{n}}}}
		\Bigg\{{1\ov 2}
			g_1 \paren{ {\epsilon_n\ov c_W} + s_Wt_W {\epsilon_nm_{Z,0}^2 \ov M_n^2-m_{Z,0}^2} }
			+{1\ov 2}  \paren{g_2 s_W {\epsilon_n m_{Z,0}^2 \ov M_n^2-m_{Z,0}^2} }
		\Bigg\},
}
with
\al{
\chi_{V^{\paren{n}}} := \paren{p_\nu - k_\nu}^2 - (m_{V^{\paren{n}}}^\tx{phys})^2
= - \paren{ 2m_e T_e + (m_{V^{\paren{n}}}^\tx{phys})^2 },
}
where that $m_{V^{\paren{n}}}^\tx{phys}$ represents the physical mass of $V^{\paren{n}}$.
\end{itemize}

Note that if we consider the mass perturbation to the 1st order, the physical masses are described as
\al{
(m_Z^\tx{phys})^2 &\to m_{Z,0}^2,&
(m_{V^{\paren{n}}}^\tx{phys})^2 &\to M_n^2,&
}
where { we} refer to Eq.~\eqref{eq:L_eff_diagonalised}.
And, for the $e$-$\nu$ scattering with the kinetic energy of the electron
$T_e$ much smaller than $m_W$ and $m_{Z,0}$,
the following approximation is valid,
\al{
\chi_W &\approx - {\sqrt{2} \paren{g_2}^2 \ov 8 G_F},&
\chi_Z &\approx - {\sqrt{2} \paren{g_2}^2 \ov 8 G_F c_W^2}.&
}
We will adopt this approximation in our calculations.

\subsection{Anomalous magnetic moment of Muon}

\begin{figure}[t]
\centering
\includegraphics[width=0.3\textwidth]{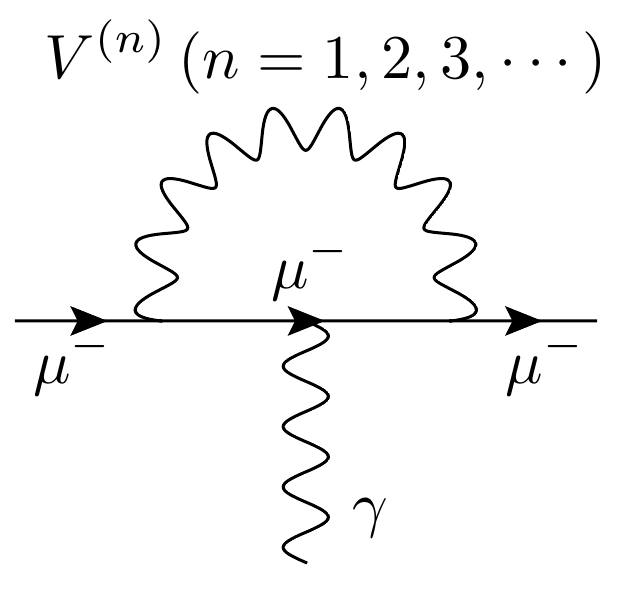}
\caption{
The topology of the Feynman diagram describing additional loop contributions of the $U\fn{1}_{L_\mu - L_\tau}$ KK gauge bosons
to the anomalous magnetic moment of the muon is shown.
}
\label{fig:muon_g-2_diagram}
\end{figure}

Each of the $U\fn{1}_{L_\mu - L_\tau}$ KK gauge bosons contributes to the anomalous magnetic moment of the muon,
parametrised as $a_\mu := \paren{g-2}_\mu/2$, through the Feynman diagram shown in Fig.~\ref{fig:muon_g-2_diagram}. According to Ref.~\cite{Leveille:1977rc} (see also \cite{Bodas:2021fsy}), for a general (flavour-{ conserving}) $Z'$ interactions with the charged leptons, {  the interaction lagrangian is given by} :
\al{
{\cal L}
	\supset
		\sum_{\alpha=e,\mu,\tau}
		\sqbr{
			\ol{l_\alpha} \gamma^\mu \paren{C_{V\alpha} + C_{A\alpha} \gamma_5} l_\alpha {Z^{'}_{\mu}}
		},
}
{and} the contribution to $\Delta a_\mu := a_\mu^\tx{exp} - a_\mu^\tx{SM}$ can be formulated as
\al{
\Delta a_\mu
	&=
		\frac{m_\mu^2}{4\pi^2 m_{Z'}^2}
		\sqbr{
			\paren{C_{V\mu}}^2 \int_0^1 \!\!\! dx \, \frac{x^2\paren{1-x}}{1-x+x^2 m_\mu^2/m_{Z'}^2}
			-
			\paren{C_{A\mu}}^2 \int_0^1 \!\!\! dx \, \frac{x\paren{1-x}\paren{4-x}+2x^3 m_\mu^2/m_{Z'}^2}
				{1-x+x^2 m_\mu^2/m_{Z'}^2}
		},
	\label{eq:muon_g-2_formula}
}
where $m_{Z'}$ represents the mass of the $Z'$ gauge boson.
Note that the above form for $\Delta a_{\mu}$ is additive, { and thus, 
it is straightforward to consider contributions from multiple gauge bosons based 
on this expression.}
For { each $V^{\paren{n}}$, the respective contribution to $\Delta a_{\mu}$ can be estimated
using the following replacement} (refer to Eq.~\eqref{eq:S_eff-Vfree-2} and Section~\ref{sec:diff-cross-section}):
\al{
m_{Z'} &\to M_n,&
C_{V\mu} &\to \wt{c}_{V^{\paren{n}}}^{\,\mu} {\chi_{V^{\paren{n}}}},&
C_{A\mu} &\to \wt{d}_{V^{\paren{n}}}^{\,\mu} {\chi_{V^{\paren{n}}}},&
}
{where multiplying $\chi_{V^{\paren{n}}}$ results in removing the propagator part of $e^- \nu_X \to e^- \nu_X$.}

Here, we provide a brief summary of the latest anomalous $\paren{g-2}_\mu$ digits of
the experimental measurement and the theoretical prediction of the SM.
The latest world average of experimental results is announced in~\cite{Muong-2:2023cdq} (cf.~\cite{Muong-2:2006rrc,Muong-2:2021ojo})
as
\al{
a_\mu^\tx{exp} 
	=
		116 \, 592 \, 059(22) \times 10^{-11},
}
while a recent comprehensive review~\cite{Aoyama:2020ynm} provides the updated theoretical prediction,
\al{
a_\mu^\tx{SM}
	=
		116 \, 591 \, 810(43) \times 10^{-11}.
}
The above two { numbers} lead to
\al{
\Delta a_\mu
	=
		249(48) \times 10^{-11},
}
where the discrepancy corresponds to a $\sim 5.2\sigma$ confidence level.\footnote{
{Note that results with less tension were reported
based on the lattice calculations~\cite{Borsanyi:2020mff,Boccaletti:2024guq} and
the measurement of $e^- e^+ \to \pi^+ \pi^-$ cross section~\cite{CMD-3:2023alj}.
See also fits to other electroweak precision observables~\cite{Crivellin:2020zul}.}
}

\section{Details of Experiments
\label{sec:Experiments}}

In this section, we provide a brief description of the experiments { which 
have measured or will measure E$\nu$ES processes, and which we will be focusing on}.

\subsection{DUNE ND}

The DUNE experiment~\cite{DUNE:2016hlj,DUNE:2020ypp} is an upcoming leading-edge long-baseline neutrino experiment, where a near detector~(ND) will be installed  at Fermilab for precise measurement of the neutrino flux and a far detector will be operational at the Sanford Underground Research Facility.
This apparatus will be essential for studies of not only neutrino oscillations but also for measuring E$\nu$ES processes precisely.

The differential event numbers at the detector are formulated as follows (see e.g.,~\cite{Melas:2023olz,Chakraborty:2021apc}):
\al{
\frac{d N^\tx{mode}}{d T_e}
	=
		t^\tx{mode}_\tx{run} N_e N_\tx{POT}
		\sum_{X: e,\ol{e},\mu,\ol{\mu}}
		\int_{E^\tx{min}_{\nu_X}}^{E_{\nu_X}^\tx{max}}
		dE_\nu
		\frac{d \Phi^\tx{mode}_{\nu_X}\fn{E_\nu}}{d E_\nu} \sqbr{ \frac{d \sigma_{e \nu_X}\fn{T_e, E_\nu}}{d T_e} },
	\label{eq:diffN-for-Te-WO}
}
where the differential cross-section of the $e^- \nu_X \to e^- \nu_X$ scattering calculated in the laboratory frame is represented as $d\sigma_{e \nu_{X}}/dT_e$ and {acceptance and resolution effects are neglected}.
$E_\nu$ and $T_e$ are the energy of the incoming neutrino and the kinetic energy of the final-state electron, respectively, where we find
$E_e = T_e + m_e$
with the energy of the final-state electron $E_e$ and the electron's mass $m_e$.
${d \Phi^\tx{mode}_{\nu_X}}/{d E_\nu}$ is the differential neutrino flux for the $X$ flavour, where two different kinds of fluxes can be provided in the forward horn current~(FHC) mode ($\nu_\mu$-dominant mode) and the reverse horn current~(RHC) mode ($\overline{\nu_{\mu}}$-dominant mode).
$E^\tx{min}_{\nu_X}$ and $E^\tx{max}_{\nu_X}$ are the minimum and maximal energy considered in the differential flux ${d \Phi^\tx{mode}_{\nu_X}}/{d E_\nu}$, where their dependence on the type of the running modes is suppressed.
{In each running mode, we will neglect the corresponding subdominant neutrino fluxes.
Eq.~\eqref{eq:formula-Tmax_e} will be used for determining $E^\tx{min}_{\nu_X}$ and $E_{\nu_X}^\tx{max}$ will be taken sufficiently { large} to cover the major peak of the corresponding neutrino flux.}

The event number in a given energy bin $T_e \in \sqbr{T_i, T_i + \Delta T_i}$ is computed as
\al{
N_i^\tx{mode} = \int_{T_i}^{T_i + \Delta T_i} dT_e \frac{d N^\tx{mode}}{d T_e}.
}
\begin{itemize}
\item
$N_\tx{POT}$ represents the number of protons on target per year (POT meaning Proton On Target).
Under the assumption of a $120\,\text{GeV}$ proton beam  { $N_\tx{POT} = 1.1 \times 10^{21}\,\tx{(\#/\tx{year})}$}~\cite{DUNE:2016hlj}.
\item
$t^\tx{mode}_\tx{run}$ is the running time of the measurement in a mode.
\item
$N_e$ represents the total electron number of the fiducial mass of the detector.
When we consider the liquid argon time-projection chamber~(LArTPC) with a fiducial mass of argon
$M_\tx{Ar}^\tx{fiducial} = 75\,\tx{ton}$ taken in~\cite{Chakraborty:2021apc} based on Ref.~\cite{DUNE:2020ypp}.
$N_e$ can be evaluated as
\al{
N_e = \frac{18 M_\tx{Ar}^\tx{fiducial}}{m_\tx{Ar}},
}
where the factor $18$ originates from the fact that each argon atom possesses 18 electrons,
and $m_\tx{Ar} = 39.95\,\tx{u} = 37.21\,\tx{GeV}$ is the atomic mass of ${}^{40}\tx{Ar}$.
\item
We will adopt the set of estimations of the differential flux provided in Figure~1 of~\cite{Marshall:2019vdy}, where
the fluxes are provided in the unit of $\paren{\tx{GeV}\,\tx{m}^2\,\tx{POT}}^{-1}$.\footnote{
Note that another set is available as Figure~4 of~\cite{Miranda:2021mqb}, where not only the fluxes for the detector on the beam axis
but also those for three off-axis locations; the original data is stored in the server~\cite{L-Fields-original}.
Data with more varieties of off-axis locations is found at~\cite{DUNE-flux-Perez-Gonzalez}.
Another important resource is Fig.~5 of Ref.~\cite{deGouvea:2019wav}.
See also Figures~4.4 and 4.5 of~\cite{DUNE:2020ypp}.
}
\item
According to~\cite{deGouvea:2019wav},
we can consider a recoil energy threshold $T^\tx{th}_e = 50\,\tx{MeV}$ and restrict our analysis in the range
$T^\tx{th}_e \leq T_e \leq T^\tx{max}_e$ with $T^\tx{max}_e = 20\,\tx{GeV}$.
\item
{ Please refer} to Appendix~\ref{sec:kinematics} for other details of kinematics.
\end{itemize}

As mentioned in~\cite{DeRomeri:2019kic}, the main backgrounds that originate from charged-current quasielastic~(CCQE) neutrino scattering on LArTPC, $\nu_e A \to e^- A'$ and $\pi^0$ misidentifications, $\nu A \to \nu \pi^0 A$ can be vetoed by a kinematical cut on the variable $E_e \theta_e^2$~\cite{MINERvA:2015nqi,Marshall:2019vdy},
where $\theta_e$ is the scattering angle for the electron in the final state.
Thereby, in the following, we also provide the formulation in terms of the quantity $E_e \theta_e^2$.
From the kinetic relationship,
\al{
1 - \cos{\theta_e} \simeq \frac{m_e}{E_e} \paren{1-y},
	\label{eq:theta_e-kinematics}
}
where
\al{
	y := {T_e \ov E_\nu}
	\label{eq:y_inelasticity}
}
denotes the inelasticity which takes values in the range $T^\tx{th}_e/E_\nu \leq y \lesssim 1$, and we used the approximation $\ab{\pmb{k}_e} \simeq E_e$, which is valid if $E_e \gg m_e$; $\pmb{k}_e$ is the three-dimensional final-state momentum of the electron.
Since E$\nu$SS is forward peaked, particularly for an energetic incoming neutrino,  the condition $\ab{\theta_e} \ll 1$ is realised.
So from Eq.~\eqref{eq:theta_e-kinematics} we get
\al{
\frac{\theta_e^2}{2} \simeq \frac{m_e}{E_e} \paren{1-y}
\quad \Leftrightarrow \quad
T_e \simeq E_\nu \paren{1 - \frac{E_e\theta_e^2}{2m_e}},
	\label{eq:Te-to-Eethetaesq}
}
where an upper limit on $E_e\theta_e^2$ is easily read off as
\al{
E_e\theta_e^2 < 2 m_e.
	\label{eq:Eethetasq-upper}
}
It is easy to reach the formulation in terms of $E_e \theta_e^2$; refer to Eq.~\eqref{eq:diffN-for-Te-WO},
\al{
\frac{d N^\tx{mode}}{d \fn{E_e \theta_e^2}}
	=
		t^\tx{mode}_\tx{run} N_e N_\tx{POT}
		\sum_{X: e,\ol{e},\mu,\ol{\mu}}
		\int_{E^\tx{min}_{\nu_X}}^{E_{\nu_X}^\tx{max}}
		dE_\nu
		\frac{d \Phi^\tx{mode}_{\nu_X}\fn{E_\nu}}{d E_\nu} \sqbr{ \frac{d \sigma_{e \nu_X}}{d\fn{E_e \theta_e^2}} },
	\label{eq:diffN-for-Eethetaesq-WO}
}
with
\al{
\frac{d \sigma_{e \nu_X}}{d\fn{E_e \theta_e^2}}
	&=
		\frac{d \sigma_{e \nu_X}}{dT_e} \ab{ \frac{dT_e}{d\fn{E_e \theta_e^2}} } \notag \\
	&\simeq
		\frac{E_\nu}{2m_e}
		\sqbr{\frac{d \sigma_{e \nu_X}}{d T_e}}_{T_e \to E_\nu \paren{1 - \frac{E_e\theta_e^2}{2m_e}}},
}
where we used Eq.~\eqref{eq:Te-to-Eethetaesq}, and
${d \sigma_{e \nu_X}}/{d\fn{E_e \theta_e^2}}$ is a function of $E_e \theta_e^2$ and $E_\nu$, and thus
$E_e$ and $\theta_e^2$ do not appear individually.

Here, we review how to do a simplified statistical analysis for the DUNE ND experiment; 
for simplicity the acceptance and resolution effects are not taken into account.
The following discussion is based on Ref.~\cite{Chakraborty:2021apc}.
From Eq.~\eqref{eq:diffN-for-Eethetaesq-WO}, we can obtain the expected number of events $N_a^\tx{mode}$ where
the value of $E_e \theta_e^2$ in the domain $\sqbr{ \paren{E_e \theta_e^2}_a, \paren{E_e \theta_e^2}_a + \Delta\paren{E_e \theta_e^2}_a }$ is evaluated as
\al{
N_a^\tx{mode}
	:=
		\int_{\paren{E_e \theta_e^2}_a}^{\paren{E_e \theta_e^2}_a + \Delta \paren{E_e \theta_e^2}_a}
		\frac{d \sigma_{e \nu_X}}{d\fn{E_e \theta_e^2}} d\fn{{E_e \theta_e^2}},
}
where the index $a$ discriminates event bins.
Note that the domain of $E_e \theta_e^2$ is
\al{
0 \leq E_e \theta_e^2 \leq 2 m_e,
}
due to Eq.~\eqref{eq:Eethetasq-upper} and the fact that $\theta_e$ can take a zero value.
We follow the scheme of the bin-by-bin analysis (and corresponding $\chi^2$ forms) considered in Ref.~\cite{Chakraborty:2021apc}.
Here, the four bins of $E_e \theta_e^2$ are considered
as $\sqbr{0,\, 0.5 m_e}$, $\sqbr{0.5 m_e,\, m_e}$, $\sqbr{m_e,\, 1.5 m_e}$, and $\sqbr{1.5 m_e,\, 2m_e}$,
and the following $\chi^2$ variable is defined:
\al{
\chi^2
	:=
		\min_{\alpha,\beta}
		\sqbr{
			\sum_{i=1}^{4}
			\frac{ \paren{N^i_\tx{NP} - \paren{1+\alpha} N^i_\tx{SM} - \paren{1+\beta} N^i_\tx{BG}}^2 }
			       {N^i_\tx{NP}}
			+ \frac{\alpha^2}{\sigma^2} + \frac{\beta^2}{\sigma^2}
		},
}
where $N^i_\tx{SM}$, $N^i_\tx{BG}$ and $N^i_\tx{NP}$ are the corresponding event numbers
{{ in the SM only, in the backgrounds and as computed in 
our scenario (including the background events),} respectively,}
in the $i$-th bin of $E_e \theta_e^2$.
The two nuisance parameters {$\alpha$ and $\beta$ are introduced and $\sigma$ represents $5\%$ systematic uncertainties}.

The total $\chi^2$ variable is defined as those in the FHC ($\nu_\mu$-dominating) mode and in the RHC ($\overline{\nu_\mu}$-dominating) mode:
\al{
\chi^2_\tx{tot} := \chi^2_\tx{FHC} + \chi^2_\tx{RHC}.
}
As mentioned, the main backgrounds for E$\nu$ES originate from the CCQE neutrino scattering on LArTPC, $\nu_e A \to e^- A'$ and the $\pi^0$ misidentifications, $\nu A \to \nu \pi^0 A$.
The author of~\cite{deGouvea:2019wav} simulated these backgrounds using the NuWro event generator~\cite{Golan:2012wx},\footnote{
The backgrounds are dependent on the information on the neutrino fluxes.
We ignore this difference since the event number of the backgrounds is sufficiently fewer than the number of expected signal events.
}
and allowed a $10\%$ normalisation uncertainty for both of { these.
We have used} their results for our analysis.

\subsection{CHARM-II}

The CHARM-II experiment used a horn-focused $\nu_{\mu}$ and $\ol{\nu_{\mu}}$ beam produced by the Super Proton Synchrotron~(SPS) at CERN~\cite{CHARM-II:1993phx,CHARM-II:1994dzw}.
The mean energy in the $\nu_\mu$ mode is $23.7\,\tx{GeV}$,
while that of $\ol{\nu_{\mu}}$ in the $\ol{\nu_{\mu}}$ mode is $19.1\,\tx{GeV}$.
The strengths of the other neutrino beams, concretely $\ol{\nu_{\mu}},\,\nu_{e},\,\ol{\nu_{e}}$ components in the $\nu_\mu$ mode,
and ${\nu_{\mu}},\,\nu_{e},\,\ol{\nu_{e}}$ components in the $\ol{\nu_\mu}$ mode, are subleading; refer to Table~1 of~\cite{CHARM-II:1993phx}.
Between 1987 and 1990, more than $2000$ $\nu e^-$ events have been recorded in both of { these} beam modes.
In Fig.~1 of~\cite{CHARM-II:1993phx}, $4+4$ unfolded differential cross sections $d\sigma/dy$
[refer to Eqs.~\eqref{eq:diff-cross-section-1}, \eqref{eq:diff-cross-section-2} and \eqref{eq:y_inelasticity}] for $e^- \nu_\mu \to e^- \nu_\mu$ and
$e^- \ol{\nu_\mu} \to e^- \ol{\nu_\mu}$ are provided in arbitrary units; see also Table~2 of~\cite{CHARM-II:1993phx}.

\subsection{Borexino}

Borexino is a 280-ton liquid scintillator detector located at the Laboratori Nazionali del Gran Sasso
in Italy~\cite{Kumaran:2021lvv}.
This experiment measures the interaction rate of the mono-energetic $862\,\tx{keV}$ ${}^7 \tx{Be}$ solar neutrino~\cite{Bellini:2011rx}.
According to~\cite{Kamada:2018zxi}, it is possible to require that the total reaction rate derivate from the SM prediction
no more than $8\%$~\cite{Bellini:2011rx} and obtain the corresponding bound.
The { minimum} value of $T_e$ is $T_\tx{min} \simeq 270\,\tx{keV}$ for the Borexino experiment~\cite{Bellini:2011rx}.

The total reaction rate at Rorexino is given by~\cite{Kamada:2018zxi}
\al{
\frac{d R^\tx{Borexino}}{d T_e}
	&=
		a^\tx{Borexino} \times
		\int_{E^\tx{min}_{\nu_{{e}}}}^{E_{\nu_{{e}}}^\tx{max}}
		dE_\nu
		\frac{d \Phi^\tx{Borexino}_{\nu_{e}}\fn{E_\nu}}{d E_\nu}
		\sqbr{ \frac{d \sigma_{\tx{tot}}\fn{T_e, E_\nu}}{d T_e} } \\
	&=
		a^\tx{Borexino} \times
		\sqbr{ \frac{d \sigma_{\tx{tot}}\fn{T_e, E_\nu}}{d T_e} }_{E_\nu = 862\,\tx{keV}},
}
with
\al{
\frac{d \Phi^\tx{Borexino}_{\nu_{e}}\fn{E_\nu}}{d E_\nu}
	=& \
		\delta\fn{ E_\nu - 862\,\tx{keV} }, \\	
\sigma_\tx{tot}
	:=& \
		P_{ee} \, \sigma_{e^- \nu_{e}} + {{1 - P_{ee}} \ov 2 }
			\sqbr{ \sigma_{e^- \nu_{\mu}} + \sigma_{e^- \nu_{\tau}} }, \\
a^\tx{Borexino}
	:=& \
		t_\tx{exp} \rho_e
} 
with the electron neutrino survival probability $P_{ee} \simeq 0.51$~\cite{Bellini:2011rx} and $P_{e\mu} \simeq P_{e\tau}$~\cite{Coloma:2022umy};\footnote{
Note that a reconsideration for this part is found in~\cite{Gninenko:2020xys}.
}
$t_\tx{exp}$ and $\rho_e$ are the exposure time and the electron number density of the target, respectively.
We assumed that the neutrino spectrum is { mono-energetic} at $E_\nu = 862\,\tx{keV}$.

If we follow the above criterion on the bound~\cite{Kamada:2018zxi}, the following part of our model will be excluded:
\al{
\frac{R^\tx{Borexino}_\tx{our model}}{R^\tx{Borexino}_\tx{SM}} > 1.08, \quad \tx{and} \quad
\frac{R^\tx{Borexino}_\tx{our model}}{R^\tx{Borexino}_\tx{SM}} < 0.92,
} 
where ${R^\tx{Borexino}_\tx{our model}/}{R^\tx{Borexino}_\tx{SM}}$ does not depend on $a^\tx{Borexino}$ defined as
\al{
R^\tx{Borexino}
	:=
		\int_{T_\tx{min}}^{T_\tx{max}} dT_e \sqbr{\frac{d R^\tx{Borexino}}{d T_e}}_{E_\nu = 862\,\tx{keV}}.
}
{We will take $T_\tx{min}$ as $0.27 \times 10^{-3}\,\tx{GeV}$ and adopted the formula in Eq.~\eqref{eq:Tmax_e} for $T_\tx{max}$.}

\subsection{TEXONO}

The TEXONO experiment~\cite{TEXONO:2009knm} measured the $e^- \ol{\nu_e} \to e^- \ol{\nu_e}$ cross section with a CsI(Tl) scintillating
crystal detector setting near the Kuo-Sheng Nuclear Power Reactor in Taiwan (see also the review paper~\cite{Wong:2015kgl}).
Therefore, the neutrino flux is the standard reactor $\ol{\nu_e}$ flux, which peaks around $1\,\tx{MeV}$. However, in TEXONO,
events are selected in the range $3\,\tx{MeV} < T_e < 8\,\tx{MeV}$, so the low energy part ($E_\nu < 3 \,\tx{MeV}$) in the flux does not contribute to the signal.

{Their differential event rates $R$ are related to the differential cross section of $e^- \ol{\nu_e} \to e^- \ol{\nu_e}$ in the following way:
\al{
{d R\ov dT_e}\,
\paren{ \propto \frac{d N^\tx{TEXONO}}{d T_e} }
	=
		a^\tx{TEXONO} \times
		\int_{E^\tx{min}_{\nu_{\ol{e}}}}^{E_{\nu_{\ol{e}}}^\tx{max}}
		dE_\nu
		\frac{d \Phi^\tx{TEXONO}_{\nu_{\ol{e}}}\fn{E_\nu}}{d E_\nu}
		\sqbr{ \frac{d \sigma_{e \nu_{\ol{e}}}\fn{T_e, E_\nu}}{d T_e} },
}
where $a^\tx{TEXONO}$ is an overall factor.
Note that the total $\ol{\nu_e}$ spectrum at typical tractor operation is shown in Fig~3 of Ref.~\cite{TEXONO:2009knm}, and
measured event rates are announced in Fig.~16 (b) of Ref.~\cite{TEXONO:2009knm}.
The $\chi^2$ form was taken in the analysis made in Ref.~\cite{Lindner:2018kjo},
\al{
\chi^2_\tx{TEXONO}
	=
		\sum_{i: \tx{10 bins}} \paren{\frac{R^\tx{th}_i - R_i^\tx{exp}}{\Delta R_i^\tx{exp}}}^2,
}
where $R_i^\tx{th}$ and $R_i^\tx{exp}$ are the theoretical and measured event rates in the $i$th recoil-energy bin, and
$\Delta R_i^\tx{exp}$ is the corresponding experimental uncertainty.
The analysis window is taken $3$--$8\,\tx{MeV}$, which is divided into ten $0.5$-MeV bins uniformly.
We can read off $R_i^\tx{exp}$ and $\Delta R_i^\tx{exp}$ from Fig.~16 (b) of Ref.~\cite{TEXONO:2009knm}.
$R_i^\tx{th}$ can be evaluated in the following way:
\al{
R_i^\tx{th}
	=
		\int_{E_\tx{$i$th bin}^\tx{min}}^{E_\tx{$i$th bin}^\tx{max}}
		d T_e
		\sqbr{ \frac{d R^\tx{TEXONO}}{d T_e} },
}
where you should be careful that, for each bin, the corresponding lower end of the $E_\nu$ integral, $E_{\nu_{\ol{e}}}^\tx{min}$,
(inside $dR^\tx{TEXONO}/dT_e$)
should be set appropriately by use of the formula in Eq.~\eqref{eq:formula-Tmax_e}.\footnote{
Note that, e.g., for the 1st bin ($i=1$), the parameters take
\al{
E_\tx{1st bin}^\tx{min} &= 3\,\tx{MeV},&
E_\tx{1st bin}^\tx{max} &= 3.5\,\tx{MeV},&
E_{\nu_{\ol{e}}}^\tx{min} &\to 3.24\,\tx{MeV}.
}
}
{$E_{\nu_{\ol{e}}}^\tx{max}$ will be taken as $\simeq 7 \times 10^{-3}\,\tx{GeV}$.}
Note that $a^\tx{TEXONO}$ can be calibrated using the SM differential cross section.

\subsection{GEMMA}

A major target of the GEMMA experiment~\cite{Beda:2009kx,Beda:2010hk} is measuring the neutrino magnetic moment by a High Purity Germanium~(HPGe) detector setting near the Kalinin Nuclear Power Plant in Russia.
The information on the $e^- \ol{\nu_e} \to e^- \ol{\nu_e}$ elastic scattering can also be extracted.
Reduction of the SM background is realised by focusing only on very low recoil energy events, from 3 keV to 25 keV.
Due to the low-energy nature of events, this data may not be very effective in putting constraints on MeV-scale physics.
Therefore, we skip the analysis of the resultant data of this experiment.

\subsection{{$\rho$ parameter}}

For the sake of rigour and safety, the constraint imposed by electroweak precision measurements is considered here.
The tree-level constraint via the electroweak precision measurements is described by the $\rho$ parameter defined as
\al{
\rho_0
	:=
		{ m_W^2 \ov m_Z^2 c_W^2},
}
where it should take $1.0000 \pm 0.0005$ (via the experimental constraints on the Peskin-Takeuchi $S$ and $T$ parameter taking $S = 0.05 \pm 0.07$ and $T = 0.00 \pm 0.06$)~\cite{ParticleDataGroup:2024cfk}.
If we consider the second-order mass perturbation, as shown in Eq.~\eqref{eq:masssq-perturbed}, this variable is estimated as
\al{
\rho_0
	=
		{ m_W^2 \ov m_{Z,0}^2 c_W^2  \sqbr{ 1 - t_W^2 m_{Z,0}^2 \sum_n {\epsilon_n^2 \ov M_n^2 - m_{Z,0}^2} }}
	\simeq
		\sqbr{ 1 + t_W^2 m_{Z,0}^2 \sum_n {\epsilon_n^2 \ov M_n^2 - m_{Z,0}^2} } =: 1 + \delta \rho_0,
}
where we assumed that $\epsilon_n^2$ is sufficiently small.

We will estimate $\delta \rho_0$ in the flat background under the simplification $\epsilon_n \to \epsilon$, where we ignore the dependence on the KK mode, and we obtain
\al{
\delta \rho_0 \sim t_W^2 \epsilon^2 {\pi m_{Z,0} \ov {8} M_\tx{KK}} \tan\sqbr{ m_{Z,0} \pi \ov m_\tx{KK} }.
}
Here, if $m_{Z,0}/m_\tx{KK} \gg 1$, the tangent function fluctuates violently, and thus so we take a reasonable average as
$\tan\sqbr{ m_{Z,0} \pi \ov m_\tx{KK} } \to 1$.
So, we obtain the $2\sigma$ bound on $\epsilon$ via the PDG result of $\delta \rho_0$,
\al{
\epsilon \lesssim10^{-3} \sqrt{ M_\tx{KK} \ov 10\,\tx{MeV} } \quad \paren{2\sigma\tx{-allowed}}.
}
If we remember
\al{
\epsilon \simeq \epsilon_4 \times \paren{\tx{Wave function of a KK gauge boson at $y=y_\tx{SM}$}}
}
and the property that the value of the wave function at a point should be around unity or less due to the normalisation,
we can reach
\al{
\epsilon_4 \lesssim 10^{-3} \sqrt{ M_\tx{KK} \ov 10\,\tx{MeV} } \quad \paren{2\sigma\tx{-allowed}}.
}
Note that the estimation for the warped case is similar to the above for the flat case within the current precision of simplifying some factors.

\section{Results
\label{sec:Results}}

\begin{figure}[t]
\centering
\includegraphics[width=0.98\textwidth]{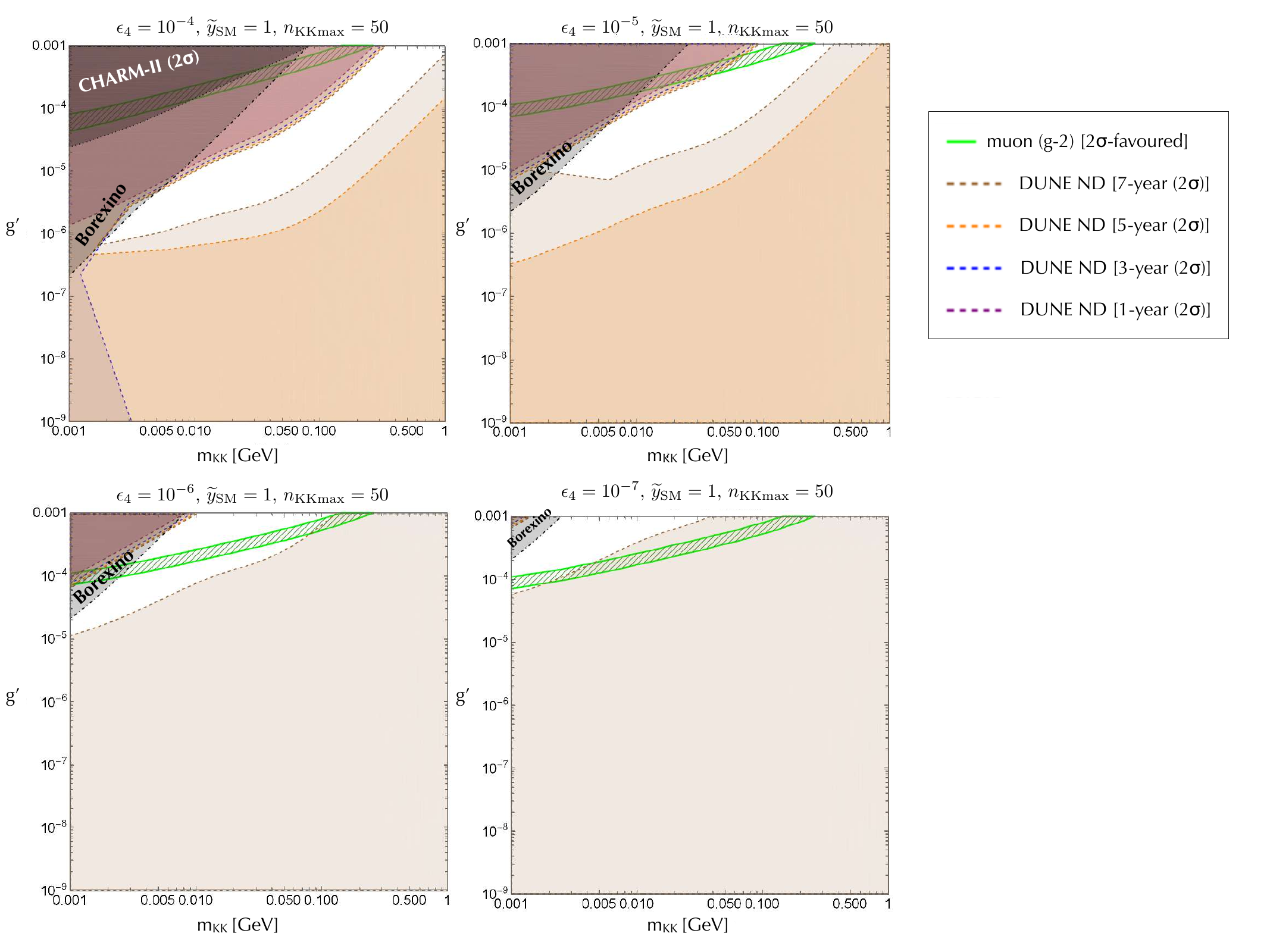}
\caption{
Current constraints (shaded by black colours) and future prospects (see the convention) in DUNE ND through E$\nu$ES processes for the four configurations with different values of $\epsilon_4$ under {$\wt{y}_\tx{SM} = 1$} when the background is flat with $n_\tx{KKmax} = 50$.
The 2$\sigma$-favoured regions of $\paren{g-2}_\mu$ are shown as the green-hatched ones.
{The $x$-year data accumulation of DUNE ND consists of $x/2$-year operations in the FHC mode and RHC mode.
Note that no 2$\sigma$ bounds are imposed from the TEXONO experimental data {and the $\rho_0$ parameter} on the shown ranges of the parameter space.}
}
\label{fig:result-flat}
\end{figure}

We will discuss current constraints and future prospects of our 5D vanilla $U\fn{1}_{L_\mu - L_\tau}$ scenario, considering it in the flat and warped backgrounds, in DUNE ND through the E$\nu$ES processes.
First, we remind the independent (effective) parameters related to E$\nu$ES (refer to Section~\ref{sec:effective-parameters}):
\al{
g', m_\tx{KK}, \epsilon_4, y_\tx{SM}, \sqbr{k \,\paren{\tx{only in warped background}}}.
}
Here, we introduce the dimensionless counterparts of $y_\tx{SM}$ and $k$ based on {a typical scale of the KK states, $R^{-1}$}, for convenience,
\al{
\wt{y}_\tx{SM} &:= {y_\tx{SM}/R},&
\wt{k} &:= {k R}.&
}

In addition, there is a practical parameter $n_\tx{KKmax}$ associated with numerical analyses, which describes
(in order of lightness) how many KK particles' effects {have been considered} in the calculation.
Literally, there are an infinite number of particles in the effective theory.
Still, as the mass increases, the effects of those particles decouple,
so for practical purposes, the value of $n_\tx{KKmax}$ should be sufficiently large for the saturation of calculations.\footnote{
Due to the existence of the factors $\epsilon_n$ and $f_n$ depending on the KK number, it is difficult to derive analytical forms of differential cross sections.
}

We keep in mind that { there are interference effects among} different intermediate states, namely,
the $W$ and $Z$ bosons, and $V^{\paren{1}}$, $V^{\paren{2}}$, $\cdots$, $V^{\paren{n_\tx{KKmax}}}$.
In our focused situation where $m_\tx{KK}$ is { in the  MeV scale, the} typical mass scale of new massive gauge bosons is sufficiently lower than the electroweak scale.
Thus, it is inadequate to consider only the 
first few light KK modes to accurately estimate the relevant cross-sections for the
numerical analyses.
On the other hand, with the issue of $\paren{g-2}_\mu$ in mind,
the parameters that describe the strength of the interactions $\br{ g', \epsilon_4 }$ tend to be ${\cal O}\fn{10^{-4}}$ to ${\cal O}\fn{10^{-3}}$.
We have checked that, if $m_\tx{KK}$ is { set to} MeV scale, and $g'$, $\epsilon_4$ are of order ${\cal O}\fn{10^{-4}}$ to ${\cal O}\fn{10^{-3}}$ or smaller,
it tends to be sufficient to take $n_\tx{KKmax} = 50$;
we discuss the status of the convergence in the supplemental Figure~\ref{fig:convergence} in Appendix~\ref{sec:supplemental-plots}.
We have checked that $n_\tx{KKmax} = 50$ suffices for all of the parameter points shown below.

\begin{figure}[t]
\centering
\includegraphics[width=0.98\textwidth]{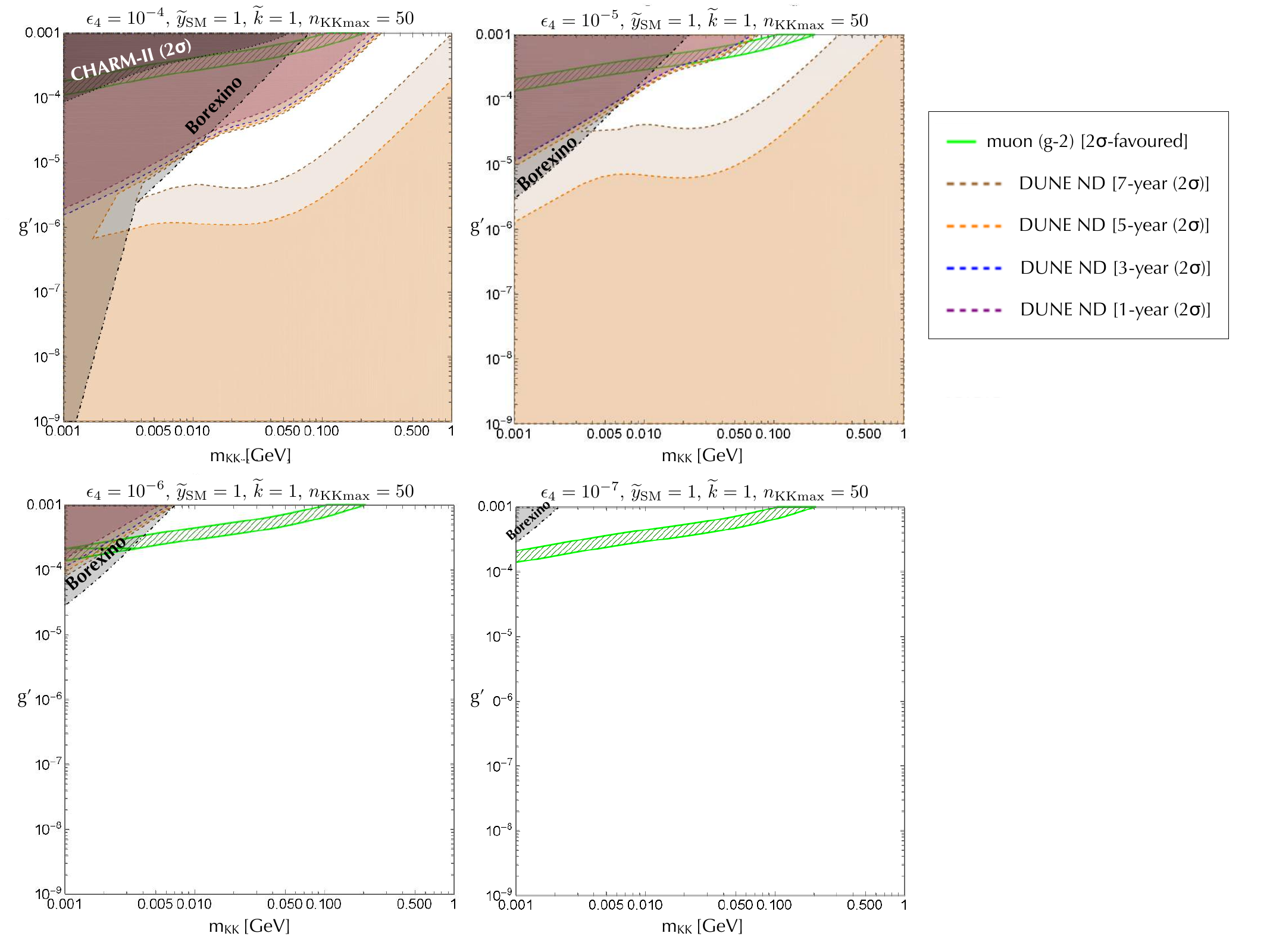}
\caption{
Current constraints and future prospects in DUNE ND through E$\nu$ES processes for the four configurations with different values of $\epsilon_4$ { and with} {$\wt{y}_\tx{SM} = 1 = \wt{k}$} when { the warped background is considered}.
The conventions and the { methodology} of the analysis are the same as in Fig.~\ref{fig:result-flat}.
}
\label{fig:result-warped}
\end{figure}

Figures~\ref{fig:result-flat} and \ref{fig:result-warped} provide concise summaries of the flat and warped cases, respectively,
when the {two massive parameters other than $m_\tx{KK}$} take a uniform value {($\wt{y}_\tx{SM} = 1 = \wt{k}$)} for simplicity.
This simplification is not expected to compromise the essential aspects of the current scenario,\footnote{
{We provide a supplemental plot for the flat case with $\epsilon_4 = 10^{-4}$ and  $n_\tx{KKmax} = 50$,
{ while setting $\wt{y}_\tx{SM}$ to} $0$, as Figure~\ref{fig:y-zero_flat} in Appendix~\ref{sec:supplemental-plots}.
{ As} expected, no drastic difference is observed between the corresponding $\wt{y}_\tx{SM} = 1$ and $\wt{y}_\tx{SM} = 0$ cases.}
}
{while we will provide additional considerations for varying $\wt{y}_\tx{SM}$ or $\wt{k}$ in Appendix~\ref{sec:varying}.}
It is useful to remember the property of the current setup that the KK gauge bosons do not directly couple with the electron;\footnote{
{So, the bound of TEXONO does not depend on $g'$.
The null direct coupling to the electron of the new gauge bosons and the smallness of the focused values of $\epsilon_4\,\paren{=10^{-4},10^{-5},10^{-6},10^{-7}}$
can explain the observation that there is no $2\sigma$-bound on the focused configurations in Figs.~\ref{fig:result-flat} and \ref{fig:result-warped}.}
}
they can interact with the electron through the mixings with the $Z$ boson.
So, if $\epsilon_4$ is set to be zero, the { additional contribution to the E$\nu$ES cross sections get to zero}, while even though $g'$ is minuscule,
the new physics contribution to the scattering cross sections is not necessarily smaller.

The first significant issue we can read from Figs.~\ref{fig:result-flat} and \ref{fig:result-warped} is that the expected discovery potential of DUNE ND is very high, in particular after the accumulation of their 5-year or 7-year data is completed.
If $\epsilon_4$ is quite sizeable as ${\cal O}\fn{10^{-4}}$ to ${\cal O}\fn{10^{-3}}$, not only the unexplored $\paren{g-2}_\mu$-favoured regions,
but also regions with minuscule-$g'$  can be surveyed extensively by E$\nu$ES processes} both for the flat and warped cases.

We find the second notable point: the interference effects play a significant role in the E$\nu$ES measurement at DUNE ND.
As shown in Figs.~\ref{fig:result-flat} and \ref{fig:result-warped}, in each panel (more precisely in six out of the eight panels), a `vacant zone' is predicted 
which is insensitive to future experimental results from DUNE ND.
The origin of such zones is that the complex interference effects 
among the Standard Model contributions and contributions from the new physics, 
which can affect the relevant amplitudes both constructively and destructively to a similar extent, resulting in rather small deviations from the theoretical prediction
of the SM.
Another interesting point is that when the background is flat, even if the $\epsilon_4$ parameter is about as small as
${\cal O}\fn{10^{-7}}$ to ${\cal O}\fn{10^{-6}}$, interference effects still remain significant and many parameter regions can be surveyed in a 7-year run of DUNE ND.
More data are needed to survey the corresponding parameter regions in DUNE ND for the warped background;
{we provide the status of the 9-year as the supplemental Figure~\ref{fig:DUNE_9-year} in Appendix~\ref{sec:supplemental-plots}}.

{We also comment on the favoured magnitudes of the $U\fn{1}_{L_\mu - L_\tau}$ gauge coupling $g'$ for $\paren{g-2}_\mu$ in the current 5D setup and the original four-dimensional vanilla scenario.
Compared with Figure~4 of~\cite{Andreev:2024lps}, our 5D scenario tends to require larger coupling constants to account for $\paren{g-2}_\mu$.
This may be explained by the fact that {the effective interaction of the 
additional gauge bosons with the muon is not purely vectorlike (in the mass eigenstates considering the kinetic mixings)} and, therefore, also gives a negative contribution to $\paren{g-2}_\mu$ [see Eqs.~\eqref{eq:S_eff-Vfree-2} and \eqref{eq:muon_g-2_formula}].}
{However, for very small kinetic mixing $\epsilon_4 \ll g'$, the vector contribution dominates in the effective interactions between muon and the additional gauge bosons, and  $\paren{g-2}_\mu$ can 
be improved to match the experimental results for smaller values of $g'$ as 
compared to the four-dimensional vanilla scenario.}

\section{Conclusions
\label{sec:Conclusions}}

In this paper, we have investigated the discovery potential of the 5D-extended minimal scenario of the $U\fn{1}_{L_\mu - L_\tau}$ gauge interaction through E$\nu$ES processes in DUNE ND, while also demonstrating the constraints from the previous experiments CHARM-II, Borexino and TEXONO.
In our setup, all of the SM particles are assumed to be confined within a zero-thickness brane (located at $y=y_\tx{SM}$) of the extra compact spatial dimension. The extra gauge symmetry is defined in the 5D, and eventually, multiple massive $U\fn{1}_{L_\mu - L_\tau}$ gauge bosons appear as KK particles.

Our analysis shows that for MeV scale KK mass parameters, significantly large parameter regions in the 5D models can be probed by DUNE ND with more than
five years of accumulated data.

This may be because multiple gauge bosons contribute to the results in our model.
Although the effect of the heavier ones is suppressed, sufficient data can be accumulated to reveal the full picture. Other observed peculiarities of our model include the presence of `blind spots' in some of the parameter spaces, where the interference effects of the intermediate states cancel out the effects of the new physics (as a whole) and which cannot be explored even with seven years of data-taking by DUNE ND.
This is a characteristic property of our model, which can have complex interference effects among the SM gauge bosons and multiple new gauge bosons with similar masses due to the nature of a KK tower.

On the other hand, our scenario has the challenge of how to explore the blind spots in this parameter space expected in DUNE ND.
Importantly, in the current scenario, the coupling between the KK gauge boson and electrons can only appear in a form that is suppressed by kinetic mixing,
while the muon couples directly with the KK bosons.
Therefore, searches for MeV mass regions through physical processes involving muon pairs in the initial and/or final state, e.g., Neutrino Trident Productions~\cite{Altmannshofer:2014pba,Altmannshofer:2019zhy,Ballett:2019xoj,Shimomura:2020tmg} and NA64-$\paren{e,\mu}$~\cite{Gninenko:2018tlp,Krnjaic:2019rsv,Sieber:2021fue,NA64:2022rme,NA64:2024klw,Andreev:2024lps}, are necessary for further exploration of our 5D scenario.\footnote{
{In addition, as discussed in~\cite{Bauer:2018onh} about the four-dimensional $U\fn{1}_{L_\mu - L_\tau}$, if the mass scale of the gauge boson is $\sim {\cal O}\fn{1}\,\tx{MeV}$ or $\gtrsim 2 \times 10^2\,\tx{MeV}$, the Big Bang Nucleosynthesis or the BaBar mono-photon searches will impose some bounds on our 5D scenario. However, our scenario involves multiple gauge bosons, so discussing specific limitations requires a precise calculation; a simple recast does not work appropriately.
}}
Moreover, it is also stimulating to see how interference effects in intermediate states work in these reaction processes.
Such aspects of our scenario will be explored in the near future with careful treatment of multiple new intermediate states (KK gauge bosons)
in the formulation of signal events and necessary numerical calculations~\cite{muon-process-survey}.

In conclusion, it may be of phenomenological interest to investigate the 
possibility of addressing some of the challenges in the SM in such MeV scale 
extra-dimensional framework.
Furthermore, further phenomenologies of the case of non-standard model sectors in five-dimensional bulk space and more theoretical considerations in the case of extra dimensions in the MeV scale should also be promoted.\footnote{
To the best of our knowledge, there do not seem to be many theoretical studies on the extra dimensions of the MeV scale, but the following previous studies can be mentioned~\cite{Jaeckel:2014eba,Aydemir:2017hyf,Anchordoqui:2020tlp,Anchordoqui:2021lmm,Antoniadis:2021mqz,Anastasopoulos:2022wob}.
}

{Finally, as a theoretical consideration for the current setup, we provide a brief comment on the cutoff scale for the KK expansion.
Since each KK gauge boson is directly coupled only to the second-generation and third-generation leptons,
the effective running coupling corresponding to each KK gauge boson exhibits a four-dimensional log running property.
Of course, each running coupling has a Landau pole at some high energy scale due to Abelian nature,
but under the current initial conditions, where $g'$ is $10^{-3}$ or less at the MeV scale, the emergence of the $U(1)_{L_\mu - L_\tau}$ Landau pole
should occur at a much higher energy scale than that for the emergence of the hypercharge $U(1)$'s Landau pole.
Thus, in the current setup, the KK expansion's effective range can be stretched (at least) up to a typical scale for the Grand Unified Theories.}

\section*{Acknowledgements}
K.N. is grateful to Kin-ya Oda for fruitful conversations.
We thank Shiv Nadar Institution of Eminence for providing us with workstations for numerical calculations.
{Furthermore, we would like to thank the anonymous referee, who made numerous useful points.}

\appendix
\section*{Appendix}

\section{Kaluza-Klein Decomposition in Warped Space
\label{sec:KK-decomposition}}

In this part, we will formulate the KK decomposition of a 5D Abelian gauge boson in a warped space.
You can refer to Ref.~\cite{Ponton:2012bi} for technical details.

\subsection{General Formalism}

We consider the KK decomposition of a five-dimensional gauge field in the bulk,
where the metric is given as\footnote{
Note that $\eta_{\mu\nu} = \eta^{\mu\nu} = \tx{diag}\fn{+1,-1,-1,-1}$; $V_y = - V^y$ and $\pal_y = -\pal^y$.
}
\al{
ds^2
	=
		g_{MN} dx^M dx^N
	=
		e^{-2A\paren{y}} \eta_{\mu\nu} dx^\mu dx^\nu - dy^2,
}
where $y \in \sqbr{0, L}$ and $A\fn{y}$ is a function of $y$.
If $A\fn{y}$ is a constant, the 5d direction is flat; otherwise, it is called as warped.
Note that the nonzero components of $g_{MN}$ are
\al{
g_{\mu\nu}
	&=
		e^{-2A} \eta_{\mu\nu},&
g_{yy}
	&=
		-1,&
g^{\mu\nu}
	&=
		e^{+2A} \eta^{\mu\nu},&
g^{yy}
	&=
		-1.&
}

The free action of a five-dimensional Abelian gauge boson $V = V\fn{x^\mu,y}$ is written down as,
\al{
S_{V}
	&:=
		S_\tx{bulk} + S_\tx{fix}, \\
S_\tx{bulk}
	&:=
		\int d^4 x \int_0^L dy \sqrt{\ab{g}}
		\br{
			-\frac{1}{4} g^{MN} g^{KL} F_{MK} F_{NL}
		}, \\
S_\tx{fix}
	&:=
		\int d^4 x \int_0^L dy
		\br{
			-\frac{1}{2\xi} \sqbr{ \pal_\mu V^\mu - \xi \pal_y \paren{ e^{-2A} V_y } }^2
		}, 
}
where $F_{MN} := \pal_M V_N - \pal_N V_M$ is the 5d field strength tensor, $\xi$ is a gauge-fixing parameter, and $\ab{g} = e^{-8A}$ is the absolute value of the determinant of $g_{MN}$.
We impose the boundary condition on $V_N$ toward the 4d Minkowski directions,
\al{
V_N\fn{ \ab{x} \to \infty, y } = 0.
	\label{eq:BC-on-A_toward-x}
}
Under the boundary condition~\eqref{eq:BC-on-A_toward-x} toward the four dimensions, the variations of $V_\mu$ and $V_y$ are derived,
\al{
\delta S_V\Big|_{V_\mu}
	&=
		\int d^4 x \int_0^L dy \,
		\delta V_\mu
		\sqbr{
			\eta^{\mu\nu} \pal^2 - \paren{1 - \frac{1}{\xi}} \pal^\mu \pal^\nu - \eta^{\mu\nu} \pal_y \paren{e^{-2A} \pal_y}
			} V_\nu \notag \\
	&\quad
		+
		\int d^4x
		\sqbr{
			\delta V_\mu
			e^{-2A}
			\paren{\pal_y V^\mu - \pal^\mu V_y}
			}_{y=0}^{y=L}, \\
\delta S_V\Big|_{V_y}
	&=
		\int d^4 x \int_0^L dy \,
		\delta V_y \, e^{-2A}
		\sqbr{
			- \pal^2 + \xi \paren{\pal_y}^2 {e^{-2A}}
			} V_y
		+ \int d^4 x
		\sqbr{
			\delta V_y \, e^{-2A}
			\paren{ \pal_\mu V^\mu - \xi \pal_y \paren{e^{-2A} V_y} }
			}_{y=0}^{y=L}.
}
Via the above forms, the equations of motion~(EOM) are derived as
\al{
\sqbr{ \eta^{\mu\nu} \pal^2 - \paren{1 - \frac{1}{\xi}} \pal^\mu \pal^\nu } V_\nu
	- \eta^{\mu\nu} \pal_y \paren{e^{-2A} \pal_y V_\nu} &= 0,
	\label{eq:5D-EOM-Amu} \\
- \pal^2 V_y + \xi {\pal_y}^2 \paren{e^{-2A} V_y} &= 0.
}
Focusing the boundary conditions toward the interval, we can identify two interesting configurations:
\al{
\paren{+} \quad \Leftrightarrow \quad V_y\Big|_\tx{a boundary} &= 0,&
	\Rightarrow \pal_y V^\mu \Big|_\tx{the same boundary} &= 0,& \\
\paren{-} \quad \Leftrightarrow \quad V^\mu\Big|_\tx{a boundary} &= 0,&
	\Rightarrow \pal_y \sqbr{ e^{-2A} V_y } \Big|_\tx{the same boundary} &= 0.&
}
Four cases are labelled as
\al{
\paren{+,+},\quad
\paren{+,-},\quad
\paren{-,+},\quad
\paren{-,-},\quad
}
where the first and second entries refer to the boundary conditions at $y=0$ and $y=L$, respectively.\footnote{
When we choose $\paren{+,+}$, a massless zero mode arises for $V_\mu$,
while we choose $\paren{-,-}$, a massless zero mode arises for $V_y$.
}

\subsection{Concrete Forms for $\paren{+,-}$}

In the following, we focus on the twisted case $\paren{+,-}$, where no physical mode remains in the KK expansion of $V_y$.
This leads us to set $\xi \to \infty$, where $V_y$ becomes zero:
\al{
S_V \Big|_{\xi \to \infty}
	&=
		\int d^4 x \int_0^L dy
		\br{
			- \frac{1}{4} F_{\mu\nu} F^{\mu\nu} + \frac{1}{2} e^{-2A} \paren{\pal_y V_\mu} \paren{\pal_y V^\mu}
		},
}
where only the physical component remains in the free action.
The KK decomposition of $V_\mu$ is introduced as,
\al{
V_{\mu}\fn{x,y}
	&=
		\frac{1}{\sqrt{L}} \sum_{n} \wh{V}^{\paren{n}}_\mu\fn{x} f^{\paren{n}}_V\fn{y},
}
where the KK summation starts from $n=0$ to $n=\infty$.
Through the 4D free EOM for a 4D gauge field (in the unitary gauge):
\al{
\br{ \sqbr{ \eta^{\mu\nu} \pal^2 - \pal^\mu \pal^\nu } + M_n^2 \eta^{\mu\nu} } \wh{V}^{\paren{n}}_\nu\fn{x} = 0,
}
and the above 5D EOM in Eq.~\eqref{eq:5D-EOM-Amu}, we can reach the differential equation for $f^{\paren{n}}_V\fn{y}$:
\al{
&
	\pal_y \sqbr{ e^{-2A} \pal_y f^{\paren{n}}_V\fn{y} } + M_n^2 f^{\paren{n}}_V\fn{y} = 0, \\
\Leftrightarrow \ &
	e^{-2A} \sqbr{  \paren{f^{\paren{n}}_V}'' - 2A' \paren{f^{\paren{n}}_V}' } + M_n^2 f^{\paren{n}}_V = 0,
	\label{eq:for-orthogonality}
}
where $M_n$ is the mass of the $n$-th KK mode.
If we take the secondary form in Eq.~\eqref{eq:for-orthogonality} and
subtract the same expression with $m \leftrightarrow n$, we can find the relationship:
\al{
\paren{ e^{-2A} \sqbr{ f^{\paren{m}}_V \paren{f^{\paren{n}}_V}' - f^{\paren{n}}_V \paren{f^{\paren{m}}_V}' } }' + \paren{M_m^2 - M_n^2} f^{\paren{m}}_V f^{\paren{n}}_V = 0.
}
Since $f^{\paren{n}}_V$ or $\paren{\pal_y f^{\paren{n}}_V}$ should vanish under the current boundary conditions, we reach
\al{
\paren{M_m^2 - M_n^2} \int_0^{L} dy f^{\paren{m}}_V f^{\paren{n}}_V = 0,
}
which means that the different KK modes are orthogonal if $M_m^2 \not= M_n^2$. 
So, we can impose the orthonormality condition on $f^{\paren{n}}_V$ (after a suitable normalisation):
\al{
\frac{1}{L} \int_{0}^{L} dy f^{\paren{m}}_V f^{\paren{n}}_V = \delta_{mn}.
}
Under the above normalisation, we can get
\al{
S_V \Big|_{\xi \to \infty}
	&=
		\int d^4 x \sum_n
		\br{
			- \frac{1}{4} \paren{\pal_\mu \wh{V}^{\paren{n}}_\nu - \pal_\nu \wh{V}^{\paren{n}}_\mu} 
				\paren{\pal^\mu \wh{V}^{{\paren{n}}\nu} - \pal^\nu \wh{V}^{{\paren{n}}\mu} }
			+ \frac{1}{2} M_n^2 \eta^{\mu\nu} \wh{V}^{\paren{n}}_\mu  \wh{V}^{\paren{n}}_\nu
		}.
}

\subsection{Orthonormal mode functions for $A\fn{y} = ky$ under $\paren{+,-}$}

Hereafter, we focus on the concrete case of $A\fn{y}$:
\al{
A\paren{y} = k y,
}
where $k$ is a constant parameter with mass dimension one.
Also, we introduce the radius of the interval:
\al{
R := {L}/{\pi}.
}

In terms of the new variable,
\al{
z := e^{ky}, \qquad dz = k e^{ky} dy,
}
the above differential equation in Eq.~\eqref{eq:for-orthogonality} for $f^{\paren{n}}_V$ can be rephrased as
\al{
\frac{d^2 f^{\paren{n}}_V}{d z^2} - \frac{1}{z} \frac{d f^{\paren{n}}_V}{dz} + \lambda_n^2 f^{\paren{n}}_V = 0,
}
with the dimensionless parameter,
\al{
\lambda_n := \frac{M_n}{k}.
}
The general solution of this 2nd-order linear differential equation is represented as
\al{
f^{\paren{n}}_V\fn{y}
	&=
		N_n z \br{ J_1\fn{\lambda_n z} + b_n Y_1\fn{\lambda_n z} } \notag \\
	&=
		N_n e^{ky} \br{ J_1\fn{\lambda_n e^{ky}} + b_n Y_1\fn{\lambda_n e^{ky}} },
}
where $N_n$ and $b_n$ are constants; $J_\alpha\fn{w}$ and $Y_\alpha\fn{w}$ denote the Bessel functions of the first and second kind, respectively.
Here, we should impose the boundary conditions:
\al{
\pal_y f^{\paren{n}}_V\fn{y}\Big|_{y = 0} &= 0,&
f^{\paren{n}}_V\fn{y}\Big|_{y = \pi R} = 0.
	\label{eq:BC_twisted}
}
The first condition of Eq.~\eqref{eq:BC_twisted} fixes the parameter $b_n$ as
\al{
b_n
	&=
		\frac
		{- \lambda_n J_0\fn{\lambda_n} -2 J_1\fn{\lambda_n} + \lambda_n J_2\fn{\lambda_n}}
		{\lambda_n Y_0\fn{\lambda_n} +2 Y_1\fn{\lambda_n} - \lambda_n Y_2\fn{\lambda_n}} \notag \\
	&=
		- \frac{J_0\fn{\lambda_n}}{Y_0\fn{\lambda_n}},
}
where we used the relation that these Bessel functions $Z_\alpha\fn{w} = J_\alpha\fn{w}$ or $Y_\alpha\fn{w}$ satifsy\footnote{
You can refer to Appendix~A of Ref.~\cite{Hosotani:2006qp} for useful formulas for Bessel functions.
}
\al{
Z_{\alpha-1}\fn{w} + Z_{\alpha+1}\fn{w} = \frac{2\alpha}{w} Z_\alpha\fn{w}.
}
After the determination of $b_n$, the second condition of Eq.~\eqref{eq:BC_twisted} leads to the form
under $N_n \not= 0$ and $e^{k\pi R} \not= 0$,
\al{
Y_0\fn{\lambda_n} J_1\fn{\lambda_n e^{k\pi R}} - J_0\fn{\lambda_n} Y_1\fn{\lambda_n e^{k\pi R}} = 0,
	\label{eq:warped_twisted-KKmass_condition}
}
which provides us with the positions of $\lambda_n \, (n = 1,2,3,\cdots)$.
The normalisation constant $N_n$ is determined as follows:\footnote{
Numerical computations are necessary to determine concrete values of $\lambda_n$ and $N_n$.
}
\al{
&
\frac{1}{\pi R} \int_0^{\pi R} dy \sqbr{ f^{\paren{n}}_V\fn{y} }^2 = \delta _{nn} = 1 \notag \\
%
%
\xrightarrow[\tx{solving for } N_n]{} \quad&
	N_n
	=
	\br{ \frac{\pi R}{ 
				\paren{
				\int_0^\pi d{y} \sqbr{
					e^{k y} \br{  J_1\fn{\lambda_n e^{ky}} 
					-  \frac{J_0\fn{\lambda_n}}{Y_0\fn{\lambda_n}} Y_1\fn{\lambda_n e^{ky}} }
				}^2
				}
			}
	}^{1/2}.
	\label{eq:Nn-form}
}

Note that the KK-mass determining condition as for $\lambda_n$ in Eq.~\eqref{eq:warped_twisted-KKmass_condition}
and the normalisation condition $N_n$ in Eq.~\eqref{eq:Nn-form} depend on the product $k R$, and $M_n$ is described as
\al{
M_n = \lambda_n k.
}
the KK scale is introduced as
\al{
m_\tx{KK} := M_1 = \frac{\lambda_1}{R} \paren{kR}.
}
Note that if $e^{k \pi R} \gg 1$, $\lambda_1 \sim e^{- k \pi R}$ (see, e.g.,~Eq.~(5.4) of \cite{Hosotani:2006qp}).
So, we can represent $m_\tx{KK}$ as
\al{
m_\tx{KK} = \lambda_1 k \sim e^{- kR \pi} k,
}
where $k$ and $kR$ are taken as independent parameters.
This form tells us that $m_\tx{KK}$ is typically (much) smaller than the mass scale $k$.\footnote{
When $k \sim 10^{19}\,\tx{GeV}$ and $kR \approx 50$, we find $m_\tx{KK} \sim 10^4 \,\tx{GeV}$~\cite{Randall:1999ee}.
This means the TeV scale is naturally induced via the Planck scale only by a moderate parameter tuning $kR \approx {\cal O}\fn{10}$.
}

\subsection{Orthonormal mode functions for $A\fn{y} = 0$ under $\paren{+,-}$}

The flat case is easy.
From the first condition of Eq.~\eqref{eq:BC_twisted}, we can find
\al{
f^{\paren{n}}_V\fn{y} \propto \cos\fn{M_n y}.
}
Imposing the second condition of Eq.~\eqref{eq:BC_twisted} leads to (refer to~\cite{Kawamura:2000ev})
\al{
M_n 
	&=
		\paren{n \,{-}\, \frac{1}{2}}\frac{1}{R}, \qquad \tx{for} \ \  n = 1,2,3,\cdots.
}
The normalised form should take
\al{
f^{\paren{n}}_V\fn{y}
	&=
		\sqrt{2} \cos\sqbr{ \paren{n \,{-}\, \frac{1}{2}}\frac{y}{R} }, \qquad \tx{for} \ \  n =1,2,3,\cdots,
}
under
\al{
\frac{1}{\pi R} \int_0^{\pi R} dy f^{\paren{m}}_V\fn{y} f^{\paren{n}}_V\fn{y} = \delta_{mn}.
}

\subsection{Relevant part of 5D action (flat case)}

The concrete form of the relevant part of the 5D action for the flat case is given as follows:
\al{
S
	&\supset
		\int \! d^4x \int_0^{\pi R} \!\! dy
		\Bigg\{
			- {1\ov4} \wh{B}_{\mu\nu} \wh{B}^{\mu\nu} - {1\ov 4} W^3_{\mu\nu} W^{3\mu\nu}
			+ { \epsilon_\tx{5D} \ov 2 c_W} \wh{B}_{\mu\nu} \wh{V}^{\mu\nu}\fn{x,y}
			+ {1\ov 2} \paren{ v g_2 W^3_\mu + v g_1 \wh{B}_\mu \ov 2}^2 \nn
	&\quad	
		+ \sum_{x=e,\mu,\tau} \ol{L'}^{\,x}_L i \gamma^\mu \sqbr{ D_\mu - i g'_\tx{5D} \wh{V}\fn{x,y}} {L'}^{\,x}_L
		+ \sum_{x=e,\mu,\tau} \ol{l'}^{\,x}_R i \gamma^\mu \sqbr{ D_\mu - i g'_\tx{5D} \wh{V}\fn{x,y}} {l'}^{\,x}_R
		\Bigg\} \delta\fn{y - y_\tx{SM}} \nn
	&\quad
		+
		\int \! d^4x \int_0^{\pi R} \!\! dy
		\Bigg\{
		- {1\ov4} \wh{V}_{MN}\fn{x,y} \wh{V}^{MN}\fn{x,y} -\frac{1}{2\xi} \sqbr{ \pal_\mu \wh{V}^\mu\fn{x,y} - \xi \pal_y \wh{V}_y\fn{x,y}  }^2
		\Bigg\},
	\label{eq:5Daction-flat}
}
where $\epsilon_\tx{5D}$ and $g'_\tx{5D}$ are five-dimensional kinetic mixing parameters and $U(1)_{L_\mu-L_\tau}$ gauge coupling.
$D_\mu$ represents a four-dimensional covariant derivative; refer to Section~\ref{sec:effective-Lagrangian} for other conventions.
Under the twisted boundary conditions, we can eliminate the 5D scalar part by taking the unitary gauge $\xi \to \infty$.
The KK decomposition,
\al{
\wh{V}\fn{x,y}
	=
		{1\ov \sqrt{{\pi R}}} \sum_n \wh{V}^{\paren{n}}_\mu\fn{x} f_V^{\paren{n}}\fn{y},
		\label{eq:KK-expansion-form}
}
we obtain
\al{
\int \! d^4x \int_0^{\pi R} \!\! dy
		\Bigg[
		- {1\ov4} \wh{V}_{MN} \wh{V}^{MN} -\frac{1}{2\xi} \sqbr{ \pal_\mu {V}^\mu - \xi \pal_y  {V}_y  }^2
		\Bigg]
		\to
		\sum_n \sqbr{ - {1\ov 4} \wh{V}^{\paren{n}}_{\mu\nu} \wh{V}^{\paren{n}\mu\nu} 
		+ {1\ov 2} M_n^2 \wh{V}^{\paren{n}}_\mu \wh{V}^{\paren{n}\mu} },
}
and find the correspondence to the effective parameters in Eqs.~\eqref{eq:effectiveLagrangian-Vfree} and \eqref{eq:effectiveLagrangian-fermion},
\al{
\epsilon_n 
	&\leftrightarrow
		 {\epsilon_\tx{5D} \ov \sqrt{{\pi R}} }  f_V^{\paren{n}}\fn{y_\tx{SM}},&
g'
	&\leftrightarrow
		{g'_\tx{5D} \ov \sqrt{{\pi R}}},&
f_n
	&\leftrightarrow
		f_V^{\paren{n}}\fn{y_\tx{SM}}.&
	\label{eq:extra-dim-parameters}
}
The final form is given as
\al{
S
	&\to
		\int \! d^4x 
		\Bigg\{
			- {1\ov4} \wh{B}_{\mu\nu} \wh{B}^{\mu\nu} - {1\ov 4} W^3_{\mu\nu} W^{3\mu\nu}
			+ \sum_n { \epsilon_n \ov 2 c_W} \wh{B}_{\mu\nu} \wh{V}^{\paren{n}\mu\nu}\fn{x,y}
			+ \sum_n \sqbr{ - {1\ov 4} \wh{V}^{\paren{n}}_{\mu\nu} \wh{V}^{\paren{n}\mu\nu} 
			+ {1\ov 2} M_n^2 \wh{V}^{\paren{n}}_\mu \wh{V}^{\paren{n}\mu} } \nn
	&\quad
		+ {1\ov 2} \paren{ v g_2 W^3_\mu + v g_1 \wh{B}_\mu \ov 2}^2
		+ \sum_{x=e,\mu,\tau} \ol{L'}^{\,x}_L i \gamma^\mu \sqbr{ D_\mu - i g' \wh{V}^{\paren{n}} f_n } {L'}^{\,x}_L
		+ \sum_{x=e,\mu,\tau} \ol{l'}^{\,x}_R i \gamma^\mu \sqbr{ D_\mu - i g' \wh{V}^{\paren{n}} f_n } {l'}^{\,x}_R
		\Bigg\},
}
which includes Eqs.~\eqref{eq:effectiveLagrangian-Vfree} and \eqref{eq:effectiveLagrangian-fermion}.

\subsection{Relevant part of 5D action (warped case)}

The concrete form of the relevant part of the 5D action for the flat case is given as follows:
\al{
S
	&\supset
		\int \! d^4x \int_0^{\pi R} \!\! dy \sqrt{ \ab{g_\tx{SM}} }
		\Bigg\{
			- {1\ov4} g_\tx{SM}^{\mu\rho} g_\tx{SM}^{\nu\sigma}
			\wh{B}_{\mu\nu} \wh{B}_{\rho\sigma} - {1\ov 4} g_\tx{SM}^{\mu\rho} g_\tx{SM}^{\nu\sigma} W^3_{\mu\nu} W^{3}_{\rho\sigma}
			+ { \epsilon_\tx{5D} \ov 2 c_W} g_\tx{SM}^{\mu\rho} g_\tx{SM}^{\nu\sigma} \wh{B}_{\mu\nu} \wh{V}_{\rho\sigma}\fn{x,y} \nn
	&\quad
		+ {1\ov 2} g_\tx{SM}^{\mu\nu} \paren{ v g_2 W^3_\mu + v g_1 \wh{B}_\mu \ov 2} \paren{ v g_2 W^3_\nu + v g_1 \wh{B}_\nu \ov 2} \nn
	&\quad	
		+ \sum_{x=e,\mu,\tau} \ol{L'}^{\,x}_L i g_\tx{SM}^{\mu\nu} \gamma_\mu \sqbr{ D_\nu - i g'_\tx{5D} \wh{V}_\nu\fn{x,y}} {L'}^{\,x}_L
		+ \sum_{x=e,\mu,\tau} \ol{l'}^{\,x}_R i g_\tx{SM}^{\mu\nu} \gamma_\mu \sqbr{ D_\nu - i g'_\tx{5D} \wh{V}_\nu\fn{x,y}} {l'}^{\,x}_R
		\Bigg\} \delta\fn{y - y_\tx{SM}} \nn
	&\quad
		+
		\int \! d^4x \int_0^{\pi R} \!\! dy \sqrt{\ab{g}}
		\Bigg\{
		- {1\ov4} g^{MK}g^{NL} \wh{V}_{MN}\fn{x,y} \wh{V}_{KL}\fn{x,y}
		-\frac{1}{2\xi} \sqbr{ \pal_\mu \wh{V}^\mu\fn{x,y} - \xi \pal_y \paren{ e^{-2A} \wh{V}_y\fn{x,y}}  }^2
		\Bigg\},
	\label{eq:5Daction-warped}
}
where the metric at the position of the SM brane is introduced,\footnote{
To get the relevant solution of the 5D Einstein equation, we should introduce 5D cosmological constant and the brane tensions at $y=0$ and $y={\pi R}$~\cite{Randall:1999ee}.
}
\al{
g_\tx{SM}^{\mu\nu}
	&=
		 e^{-2A\paren{y_\tx{SM}}} \eta_{\mu\nu},&
\sqrt{ \ab{g_\tx{SM}} }
	&=
		e^{-4A\paren{y_\tx{SM}}}.&
}
After taking the unitary gauge (under the twisted boundary conditions) and KK decompositions by use of Eqs.~\eqref{eq:KK-expansion-form} and \eqref{eq:extra-dim-parameters}, the effective action reads
\al{
S
	&\to
		\int \! d^4x 
		\Bigg\{
			- {1\ov4} \wh{B}_{\mu\nu} \wh{B}^{\mu\nu} - {1\ov 4} W^3_{\mu\nu} W^{3\mu\nu}
			+ \sum_n { \epsilon_n \ov 2 c_W} \wh{B}_{\mu\nu} \wh{V}^{\paren{n}\mu\nu}\fn{x,y} \nn
	&\quad
			+ \sum_n \sqbr{ - {1\ov 4} \wh{V}^{\paren{n}}_{\mu\nu} \wh{V}^{\paren{n}\mu\nu} 
			+ {1\ov 2} M_n^2 \wh{V}^{\paren{n}}_\mu \wh{V}^{\paren{n}\mu} }
			+ {1\ov 2} {e^{-2A\paren{y_\tx{SM}}}} 
				\paren{ v g_2 W^3_\mu + v g_1 \wh{B}_\mu \ov 2} \paren{ v g_2 W^{3\mu} + v g_1 \wh{B}^\mu \ov 2} \nn
	&\quad
		+ \sum_{x=e,\mu,\tau} {e^{-2A\paren{y_\tx{SM}}}} \,
			\ol{L'}^{\,x}_L i \gamma^\mu \sqbr{ D_\mu - i g' \wh{V}^{\paren{n}} f_n } {L'}^{\,x}_L
		+ \sum_{x=e,\mu,\tau} {e^{-2A\paren{y_\tx{SM}}}} \,
			 \ol{l'}^{\,x}_R i \gamma^\mu \sqbr{ D_\mu - i g' \wh{V}^{\paren{n}} f_n } {l'}^{\,x}_R
		\Bigg\}.
}
After the field renormalisations,
\al{
v &\to e^{+A\paren{y_\tx{SM}}} v,&
L^{'x}_L &\to e^{+A\paren{y_\tx{SM}}} L^{'x}_L,&
l^{'x}_R &\to e^{+A\paren{y_\tx{SM}}} l^{'x}_R,&
}
we get
\al{
S
	&\to
		\int \! d^4x 
		\Bigg\{
			- {1\ov4} \wh{B}_{\mu\nu} \wh{B}^{\mu\nu} - {1\ov 4} W^3_{\mu\nu} W^{3\mu\nu}
			+ \sum_n { \epsilon_n \ov 2 c_W} \wh{B}_{\mu\nu} \wh{V}^{\paren{n}\mu\nu}\fn{x,y} \nn
	&\quad
			+ \sum_n \sqbr{ - {1\ov 4} \wh{V}^{\paren{n}}_{\mu\nu} \wh{V}^{\paren{n}\mu\nu} 
			+ {1\ov 2} M_n^2 \wh{V}^{\paren{n}}_\mu \wh{V}^{\paren{n}\mu} }
			+ {1\ov 2}
				\paren{ v g_2 W^3_\mu + v g_1 \wh{B}_\mu \ov 2}^2 \nn
	&\quad
		+ \sum_{x=e,\mu,\tau} 
			\ol{L'}^{\,x}_L i \gamma^\mu \sqbr{ D_\mu - i g' \wh{V}^{\paren{n}} f_n } {L'}^{\,x}_L
		+ \sum_{x=e,\mu,\tau} 
			 \ol{l'}^{\,x}_R i \gamma^\mu \sqbr{ D_\mu - i g' \wh{V}^{\paren{n}} f_n } {l'}^{\,x}_R
		\Bigg\},
}
which includes Eqs.~\eqref{eq:effectiveLagrangian-Vfree} and \eqref{eq:effectiveLagrangian-fermion}.

\section{Kinematics of $e^- \nu_X \to e^- \nu_X$
\label{sec:kinematics}}

Here, we summarise the kinematical relationships for E$\nu$ES in the laboratory frame, where
we assign $p_\nu$, $p_e$, $k_\nu$, and $k_e$ as the initial-state neutrino and electron four-dimensional momenta and those for the final state, respectively; concretely,
\al{
\paren{p_\nu}^\mu
	&=
		\paren{ E_\nu, E_\nu \, \wh{\pmb{z}} },&
\paren{p_e}^\mu
	&=
		\paren{ m_e, \pmb{0} },& \notag \\
\paren{k_\nu}^\mu
	&=
		\paren{ E'_\nu,
			\underbrace{E'_\nu \sin{\theta_\nu} \, \wh{\pmb{x}} + E'_\nu \cos{\theta_\nu} \, \wh{\pmb{z}}}_{=\pmb{k}_\nu} },&
\paren{k_e}^\mu
	&=
		\paren{ E_e, \pmb{k}_e },&
}
where the scattering angle for the neutrino and the electron in the final state are defined as those between $\wh{\pmb{z}}$ and $\pmb{k}_\nu$ and between $\wh{\pmb{z}}$ and $\pmb{k}_e$, respectively.
We introduce the kinetic energy of the final-state election defined by
\al{
T_e := E_e - m_e.
}

The four-dimensional energy-momentum conservation tells us
\al{
\paren{p_\nu}^\mu + \paren{p_e}^\mu = \paren{k_\nu}^\mu + \paren{k_e}^\mu,
}
particularly the energy part ($\mu=0$) takes the form,
\al{
E_\nu + m_e = E'_\nu + E_e,
\quad \Leftrightarrow \quad
E_\nu = E'_\nu + T_e.
}

The following relations are useful:
\al{
& 
	\paren{p_\nu - k_e}^2 = \paren{k_\nu - p_e}^2 = m_e^2 - 2 \paren{p_\nu \cdot k_e} = m_e^2 - 2 \paren{k_\nu \cdot p_e}, \notag \\
\Leftrightarrow \ &
	 \paren{p_\nu \cdot k_e} = \paren{k_\nu \cdot p_e} = m_e E'_\nu, \\[5pt]
& 
	\paren{p_\nu - k_\nu}^2 = \paren{k_e - p_e}^2 =
	- 2 \paren{p_\nu \cdot k_\nu} = 2 m_e^2 - 2 \paren{k_e \cdot p_e}, \notag \\
\Leftrightarrow \ &
	\paren{p_\nu \cdot k_\nu} = \underbrace{\paren{k_e \cdot p_e}}_{= m_e E_e} - m_e^2
	= m_e \paren{E_e - m_e} = m_e T_e = E_\nu E'_\nu \paren{1 - \cos\theta_\nu},
}
where the relation between the final two forms leads to
\al{
T_e = \frac{E_\nu E'_\nu}{m_e} \paren{1 - \cos\theta_\nu}.
	\label{eq:T_e-form-primitive}
}
From $E'_\nu = E_\nu - T_e$ and Eq.~\eqref{eq:T_e-form-primitive}, we get
\al{
T_e
	&=
		\frac{E_\nu^2 \paren{1-\cos\theta_\nu}}{m_e + E_\nu\paren{1-\cos\theta_\nu}},&
\frac{d T_e}{d \cos\theta_\nu}
	&=
		- \frac{E_\nu^2 m_e}{ \paren{E_\nu \paren{1 - \cos\theta_\nu} + m_e}^2 }, \notag \\
E'_\nu
	&=
		E_\nu \paren{\frac{m_e + E_\nu \paren{1 - \cos\theta_\nu}}{m_e + E_\nu \paren{1 - \cos\theta_\nu}}}
		- \frac{E_\nu^2 \paren{1 - \cos\theta_\nu}}{m_e + E_\nu \paren{1 - \cos\theta_\nu}} \notag \\
	&=
		\frac{m_e E_\nu}{m_e + E_\nu \paren{1-\cos\theta_\nu}}.&
}
The following relation is useful for the differential cross-section in the laboratory frame:\footnote{
Note that Eq.~(B.23) of the latest arXiv version (v1) of Ref.~\cite{Lindner:2018kjo} has a typo.
}
\al{
\frac{d \sigma}{d T_e}
	&=
		\frac{d \sigma}{d \cos\theta_\nu} \ab{ \frac{d \cos\theta_\nu}{d T_e} } \notag \\
	&=
		\frac{ \paren{E'_\nu}^2 }{32 \pi m_e^2 E_\nu^2} \langle \ab{{\cal M}}^2 \rangle_\tx{helicity}
		 \ab{ \frac{d \cos\theta_\nu}{d T_e} } \notag \\
	&=
		\frac{ 1 }{32 \pi m_e E_\nu^2} \langle \ab{{\cal M}}^2 \rangle_\tx{helicity}.
	\label{eq:dsigma-ov-dTe}
}

The variable $y := {T_e}/{E_\nu}$
denotes the inelasticity, which takes values in the range $0 \lesssim y \lesssim 1$
or $T^\tx{th}_e/E_\nu \leq y \lesssim 1$, which correspond to without or with taking into account a recoil energy threshold for the final-state electron ($T^\tx{th}_e$) at the detector.
This variable can be used instead of $T_e$.
It is easy to understand the relation,
\al{
\frac{d \sigma}{d y}
	&=
		\frac{d \sigma}{d T_e} \underbrace{\frac{d T_e}{d y}}_{= E_\nu}
		= E_\nu \sqbr{ \frac{d \sigma}{d T_e} }_{T_e \to E_\nu y}.
}

The magnitude of $\pmb{k}_e$ is described as
\al{
\ab{\pmb{k}_e}
	&=
		\sqrt{ E_e^2 - m_e^2 }
		= \sqrt{ \paren{E_e - m_e}\paren{E_e + m_e} } \notag \\
	&=
		\sqrt{T_e\paren{T_e+2m_e}}.
}

The relation between $T_e$ and $\theta_e$ is evaluated:
\al{
\underbrace{\paren{p_\nu \cdot k_e}}_{= m_e E'_\nu = m_e \paren{E_\nu - T_e}}
	&=
		E_\nu \underbrace{E_e}_{T_e + m_e} 
		- E_\nu \underbrace{ \paren{ \wh{\pmb{z}} \cdot \pmb{k}_e }}_{= \ab{\pmb{k}_e} \cos{\theta_e}} \notag \\
	\Rightarrow \quad 
	\cos{\theta_e}
	&=
		\frac{T_e \paren{E_\nu+m_e}}{E_\nu \ab{\pmb{k}_e}}
		= \frac{T_e \paren{E_\nu+m_e}}{E_\nu  \sqrt{T_e\paren{T_e+2m_e}}  },
	\label{eq:costhetae-full}
}
where the square of the above relation leads to
\al{
T_e
	=
		\frac
		{2 m_e E_\nu^2 \cos^2{\theta_e}}
		{\paren{E_\nu+m_e}^2 - E_\nu^2 \cos^2{\theta_e}}.
	\label{eq:T_e-form}
}
\begin{itemize}
\item
Form the derivative
\al{
\frac{\pal \, T_e}{\pal \cos^2\theta_e}
	=
		\frac
		{2 m_e E_\nu^2 \paren{E_\nu + m_e}^2 }
		{\paren{\paren{E_\nu+m_e}^2 - E_\nu^2 \cos^2{\theta_e}}^2} \paren{> 0},
}
we can recognise that $T_e$ is maximised at $\theta_e = 0\,(\Leftrightarrow \cos\theta_e = 1)$.
\item
For a focused $E_\nu$, the maximal kinetic energy of $T_e$ is realised at $\theta_e = 0\,(\Leftrightarrow \cos\theta_e = 1)$ as
\al{
T^\tx{max}_e\fn{E_\nu}
	=
		\frac{2 E_\nu^2}{m_e + 2 E_\nu}.
	\label{eq:Tmax_e}
}
\item
From Eq.~\eqref{eq:T_e-form}, we get
\al{
E_\nu
	=
		\frac
		{m_e T_e + \sqrt{ m_e^2 \cos^2\theta_e \paren{T_e^2 + 2 m_e T_e} }}
		{T_e \paren{\cos^2\theta_e-1} + 2m_e\cos^2\theta_e}.
}
When the kinetic energy of the electron is maximised, the energy of the neutrino is minimised.
Thus, by taking $\theta_e = 0\,(\Leftrightarrow \cos\theta_e = 1)$, we can derive the least energy that is necessary for realising
the scattered final-state electron with the kinetic energy $T_e$,
\al{
E^\tx{min}_\nu\fn{T_e}
	&=
		\frac{1}{2} \paren{T_e + \sqrt{T_e^2 + 2 m_e T_e}} = \frac{T_e + \ab{\pmb{k}_e}}{2}, 
		\label{eq:formula-Tmax_e} \\
	&\simeq
		\begin{cases}
			\sqrt{\frac{m_eT_e}{2}} & \text{for  } T_e \ll m_e, \\
			T_e + \frac{m_e}{2} & \text{for  } T_e \gg m_e.
		\end{cases}
}
Note that the above results agree with Eq.~(5.31) of~\cite{10.1093/acprof:oso/9780198508717.001.0001}.
\item
From Eq.~\eqref{eq:costhetae-full}, if $\ab{\pmb{k}_e} \simeq E_e$, which is valid if $E_e \gg m_e$, we can get
\al{
\cos{\theta_e}
	&\simeq
		\frac{T_e \paren{E_\nu+m_e}}{E_\nu E_e},
}
which is rephrased as
\al{
1 - \cos{\theta_e} \simeq \frac{m_e}{E_e} \paren{1-y}.
}
\item
Via Eq.~\eqref{eq:formula-Tmax_e} and the approximated relation in Eq.~\eqref{eq:Te-to-Eethetaesq},
\als{
T_e \simeq E_\nu \paren{1 - \frac{E_e\theta_e^2}{2m_e}},
}
we find the relation for a nonzero $\sqbr{E_e\theta_e^2}_\tx{max}$ of a bin,
\al{
E^\tx{min}_\nu
	&\simeq
		{m_e \paren{2 m_e - \sqbr{E_e\theta_e^2}_\tx{max} } \ov  2 \sqbr{E_e\theta_e^2}_\tx{max}}.
}
\end{itemize}

\section{Supplemental Plots
\label{sec:supplemental-plots}}

Here, we provide some supplemental plots that are helpful for a confirmed understanding of our analysis.

\begin{figure}[H]
\centering
\includegraphics[width=0.95\textwidth]{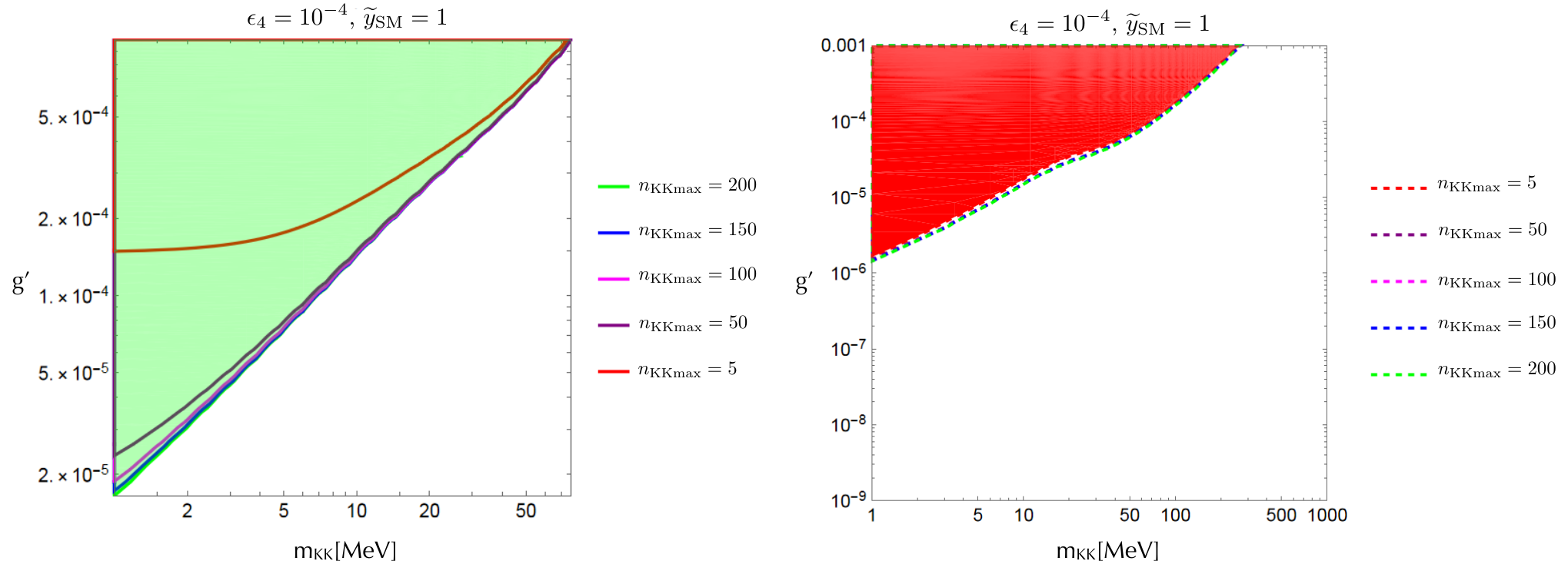}
\caption{
For the flat case with $\epsilon_4 = 10^{-4}$ and $\wt{y} = 1$,
the statuses of the convergence of the KK summation in the analyses of CHARM-II (left) and a one-year running of DUNE ND (right) are summarised
as showing the differences in the $2\sigma$ bounds.
}
\label{fig:convergence}
\end{figure}

\begin{figure}[H]
\centering
\includegraphics[width=0.65\textwidth]{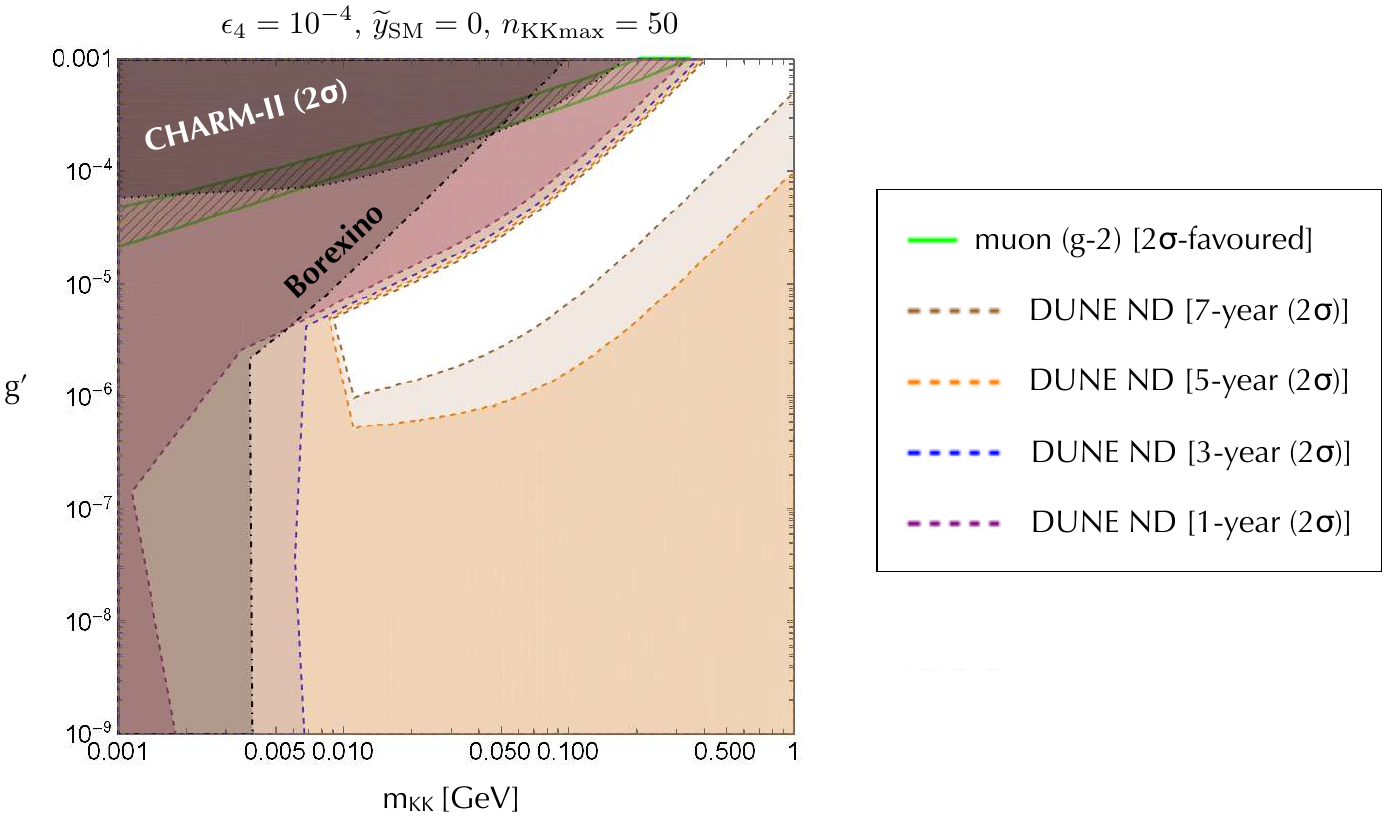}
\caption{
A supplemental plot for the flat case with $\epsilon_4 = 10^{-4}$ and  $n_\tx{KKmax} = 50$,
but $\wt{y}_\tx{SM}$ taking $\wt{y}_\tx{SM} = 0$, different from $\wt{y}_\tx{SM} = 1$ in Fig.~\ref{fig:result-flat}.
The conventions and the basic way of the analysis are the same as in Fig.~\ref{fig:result-flat}.
}
\label{fig:y-zero_flat}
\end{figure}

\begin{figure}[H]
\centering
\includegraphics[width=0.6\textwidth]{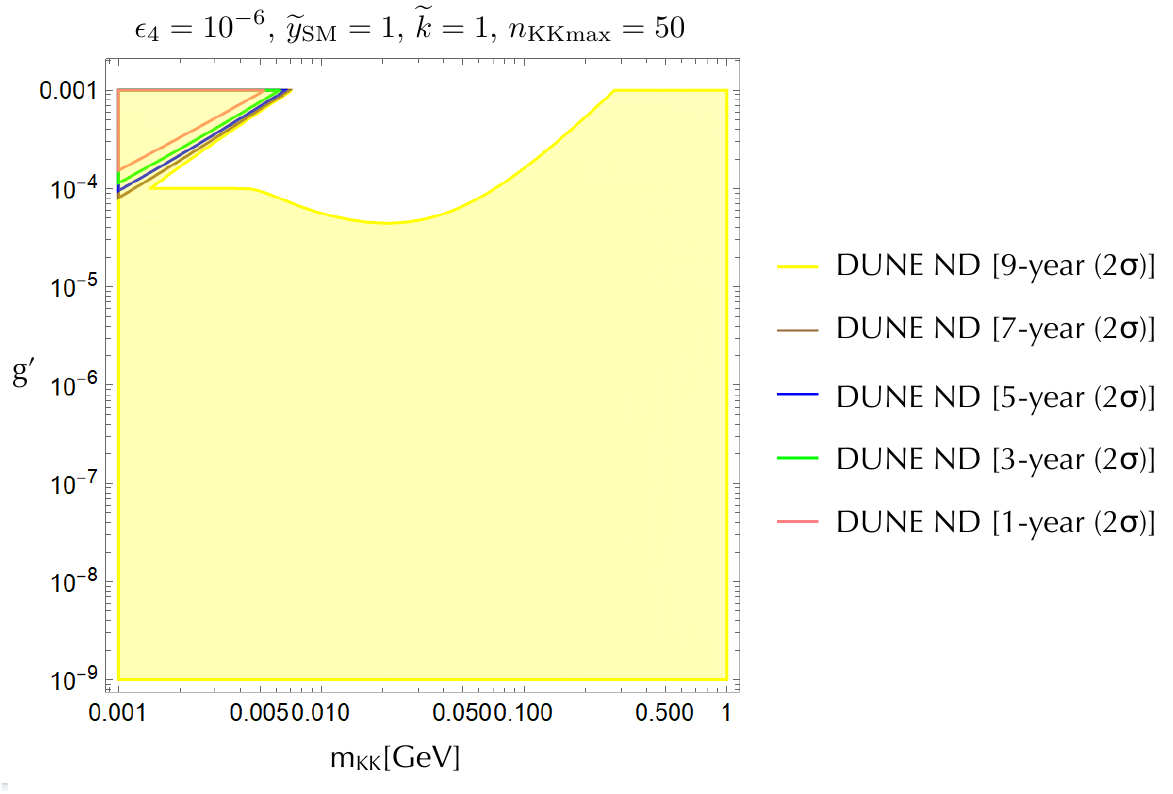}
\caption{
The expected $2\sigma$-disfavoured region of the $9$-year running of DUNE ND is shown for the warped case with $\epsilon_4 = 10^{-6}$,
$\wt{y} = \wt{k} = 1$ and $n_\tx{KKmax} = 50$,
where a `vacant zone' is observed as in the four plots of Fig.~\ref{fig:result-flat} and the two plots (for $\epsilon_4 = 10^{-4}$, $10^{-5}$)
of Fig.~\ref{fig:result-warped}.
}
\label{fig:DUNE_9-year}
\end{figure}

\section{{Varying $\wt{k}$ and $\wt{y}_\tx{SM}$ from unity}
\label{sec:varying}}

In this part, we will provide additional consideration for varying $\wt{k}$ and $\wt{y}_\tx{SM}$ from unity
in discussing current constraints and future prospects of our 5D vanilla $U\fn{1}_{L_\mu - L_\tau}$ scenario.

\begin{figure}[H]
\centering
\includegraphics[width=0.6\textwidth]{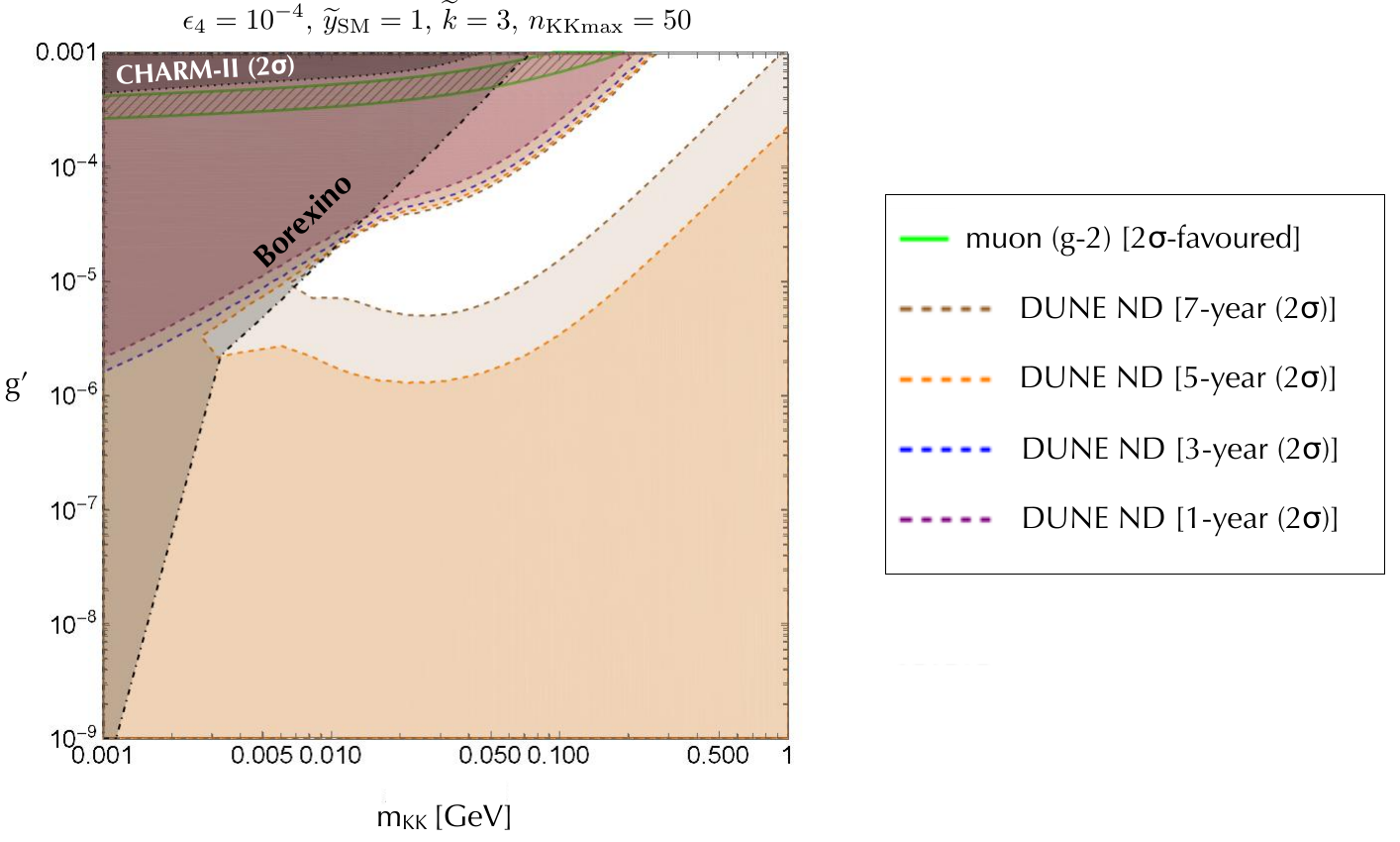}
\caption{
An additional plot for the warped case with $\epsilon_4  = 10^{-4}$, $\wt{y}_\tx{SM} = 1$, $n_\tx{KKmax} = 50$,
but $\wt{k} = 3$, different from $\wt{k} = 1$ in Fig.~\ref{fig:result-warped}.
The conventions and the basic way of the analysis are the same as in Fig.~\ref{fig:result-warped}.
}
\label{fig:case_ktilde-3}
\end{figure}

First, we will show the warped-case result for $\wt{k} = 3$ with
$\epsilon_4  = 10^{-4}$, $\wt{y}_\tx{SM} = 1$ and $n_\tx{KKmax} = 50$ in Fig.~\ref{fig:case_ktilde-3},
and compare it with the upper left panel of Fig.~\ref{fig:result-warped},
where the value of $\wt{k}$ is taken as unity instead.
One of the most distinctive differences here is the stronger 5- and 7-year limit forecast for DUNE ND.
This can be interpreted as a result of the increased value of the warp factor $\wt{k}$, which has resulted in lighter masses in KK modes
and a stronger effect of the new physics.

Second, we will consider changing $\wt{y}_\tx{SM}$ to $3$ from unity in the flat and warped cases
for $\epsilon_4  = 10^{-4}$, $\wt{y}_\tx{SM} = 1$; and also $\wt{k} = 1$ (for the warped case).
Since the Dirichlet boundary condition is imposed for the endpoint $\wt{y} = \pi$ (refer to Fig.~\ref{fig:5D-setup}),
the KK wavefunctions at the SM brane
$f_n = f_V^{\paren{n}}\fn{\wt{y}_\tx{SM}}$ vanish at $\wt{y}_\tx{SM} = \pi$ and at this time, contributions from the new physics obviously become zero.
Our choice of $\wt{y}_\tx{SM} \,(=3)$ is close to this case ($\wt{y}_\tx{SM} = \pi$), but nonetheless,
the current constraints and future prospects for the flat and warped cases are not so different from the corresponding plots,
which are the upper-left panels of Figs~\ref{fig:result-flat} and \ref{fig:result-warped}, respectively.
This observation can be understood as follows.
As shown in Fig.~\ref{fig:wavefunction-profiles}, for both the flat and warped cases,
the profiles of the first KK mode ($n=1$) tend to be small for $\wt{y}_\tx{SM} = 3$,
but those of higher modes are not suppressed in general.
Since we focus on the situation where $m_\text{KK}$ is a MeV scale, the contribution of higher modes is sizable.
Based on these observations and considerations, we can conclude that
the current constraints and future prospects are not shrunk significantly even if we place the parameter $\wt{y}_\tx{SM}$ near the Dirichlet boundary at $\wt{y} = \pi$.

\begin{figure}[H]
\centering
\includegraphics[width=0.98\textwidth]{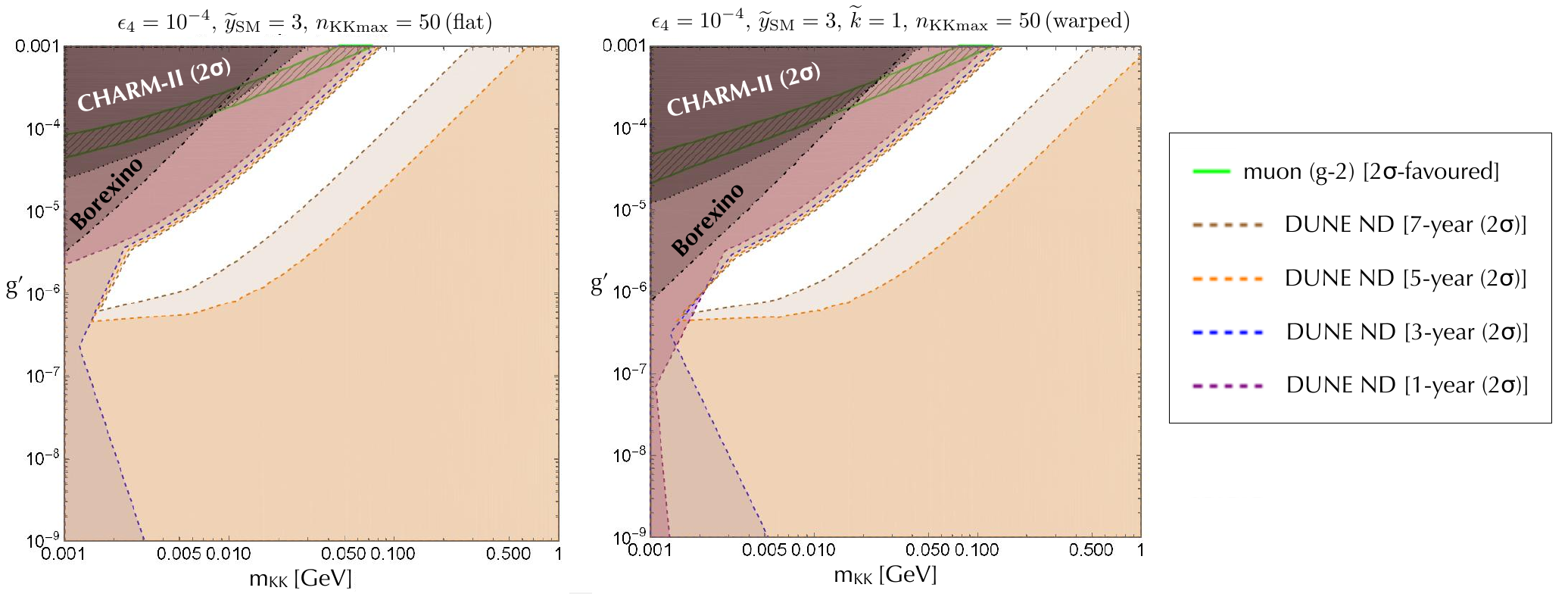}
\caption{
An additional plot for the flat (Left panel) and warped (Right panel) cases with $\epsilon_4  = 10^{-4}$, $\wt{k} = 1$ (only for the warped case), $n_\tx{KKmax} = 50$,
but $\wt{y}_\tx{SM} = 3$, different from $\wt{y}_\tx{SM} = 1$ in Figs.~\ref{fig:result-flat} and \ref{fig:result-warped}.
The conventions and the basic way of the analysis are the same as in Figs.~\ref{fig:result-flat} and \ref{fig:result-warped}.
}
\label{fig:case_ytilde-3}
\end{figure}

\begin{figure}[H]
\centering
\includegraphics[width=0.98\textwidth]{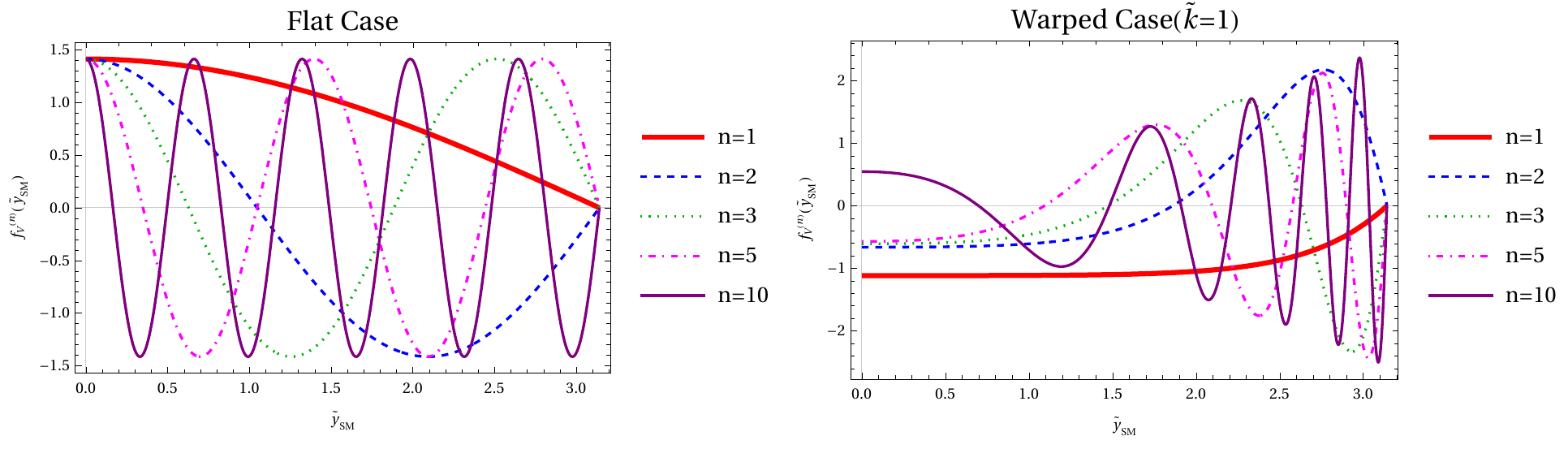}
\caption{
The wavefunction profiles $f_V^{\paren{n}}\fn{\wt{y}_\tx{SM}}$ at the SM brane $\wt{y}_\tx{SM}$ located at $\sqbr{0,\pi}$
of the five KK modes $\paren{n = 1,2,3,5,10}$ for the flat case (Left panel) and the warped case with $\wt{k} = 1$ (Right panel).
}
\label{fig:wavefunction-profiles}
\end{figure}

\bibliographystyle{utphys}
\bibliography{LmuLtau-ref,Exdim-Zprime,updated-ref}

\end{document}